\documentclass[twocolumn,showpacs,superscriptaddress,%
aps,prd,notitlepage,longbibliography,nofootinbib]{revtex4-1}

\usepackage{xcolor}
\usepackage{graphicx}
\usepackage{amsmath}
\usepackage{amsfonts}
\usepackage{amssymb}
\usepackage{maybemath}
\usepackage{bbm}
\usepackage{hyperref}
\usepackage{seceqn.arabic}

\definecolor{garrosgreen}{rgb}{0.1, 0.4, 0.1}
\definecolor{dartmouthgreen}{rgb}{0.05, 0.5, 0.06}
\definecolor{duelferred}{rgb}{0.7, 0.2, 0.1}
\definecolor{camblue}{rgb}{0.1, 0.3, 1.0}
\definecolor{oxfordblue}{rgb}{0.05, 0.2, 0.7}

\newcommand{\calD}{{\mathcal D}}
\newcommand{\calG}{{\mathcal G}}
\newcommand{\calJ}{{\mathcal J}}
\newcommand{\calO}{{\mathcal O}}
\newcommand{\calQ}{{\mathcal Q}}

\def\calA{{\mathcal A}}
\def\calB{{\mathcal B}}
\def\calC{{\mathcal C}}
\def\calD{{\mathcal D}}
\def\calE{{\mathcal E}}
\def\calF{{\mathcal F}}

\def\calJ{{\mathcal J}}
\def\calK{{\mathcal K}}

\def\calN{{\mathcal N}}
\def\calO{{\mathcal O}}
\def\calR{{\mathcal R}}
\def\calS{{\mathcal S}}
\def\calT{{\mathcal T}}
\def\calU{{\mathcal U}}
\def\calW{{\mathcal W}}

\def\calZ{{\mathcal Z}}

\def\bfM{{\bf M}}

\def\bfG{{\bf G}}

\def\bfDelta{{\bf \Delta}}

\def\sfG{{\sf G}}

\newcommand{\undq}{{\underline q}}
\newcommand{\undu}{{\underline u}}

\newcommand{\undchi}{{\underline \chi}}
\newcommand{\undJ}{{\underline J}}
\newcommand{\undzero}{{\underline 0}}

\newcommand{\cl}{{\mathrm{cl}}}
\newcommand{\LL}{{\mathrm{L}}}
\newcommand{\LLi}{{\mathrm{Li}}}
\newcommand{\TT}{{\mathrm{T}}}

\newcommand{\dd}{{\mathrm{d}}}

\newcommand{\rmT}{\mathrm{T}}

\def\Im{\mathop{\rm Im}\nolimits}

\def\tfrac12{{\textstyle\frac12}}

\def\det{\mathop{\rm det}\nolimits}

\def\dd{{\mathrm d}}
\def\ii{{\mathrm i}}
\def\ee{{\mathrm e}}
\def\cl{{\mathrm{cl}}}

\def\tfrac#1#2{ {\textstyle{\frac{#1}{#2}} } }
\def\und#1{\underline{ #1 }}

\newcommand{\intbeta}{\int_{-\beta/2}^{\beta/2}}

\definecolor{garrosgreen}{rgb}{0.1, 0.4, 0.1}
\definecolor{dartmouthgreen}{rgb}{0.05, 0.5, 0.06}
\definecolor{duelferred}{rgb}{0.7, 0.2, 0.1}
\definecolor{cambridgeblue}{rgb}{0.1, 0.3, 1.0}
\definecolor{oxfordblue}{rgb}{0.05, 0.2, 0.7}

\usepackage{color}

\begin{document}

\title{Two--Loop Corrections to the Large--Order Behavior of Correlation Functions\\
in the One--Dimensional 
\texorpdfstring{$\bm{N}$}{N}--Vector Model}

\newcommand{\addrNordita}{Nordita, Royal Institute of Technology and Stockholm University, Stockholm 106 91, Sweden}

\newcommand{\addrRolla}{Department of Physics, Missouri University of Science
and Technology, Rolla, Missouri 65409, USA}

\newcommand{\addrMTADE}{MTA--DE Particle Physics Research Group,
P.O.~Box 51, H--4001 Debrecen, Hungary}

\newcommand{\addrMilan}{Bocconi Institute for Data Science and Analytics, Bocconi University, Milano 20136, Italy}

\newcommand{\addrRomaISC}{ISC-CNR, UOS Rome, 
Universit\`{a} ``Sapienza,'' Piazzale A. Moro 2, I-00185 Rome, Italy}

\newcommand{\addrRomaSAP}{Dipartimento di Fisica, 
Universit\`{a} ``Sapienza,'' Piazzale A. Moro 2, I-00185, Rome, Italy}

\newcommand{\addrSACLAY}{IRFU/CEA,
Paris--Saclay, 91191 Gif-Sur-Yvette, France}

\author{L. T. Giorgini}
\email{email: ludovico.giorgini@su.se (author contributed equally to the submitted work). }
\affiliation{\addrNordita}

\author{U. D. Jentschura}
\email{email: ulj@mst.edu (author contributed equally to the submitted work)}
\affiliation{\addrRolla}
\affiliation{\addrMTADE}

\author{E. M. Malatesta}
\affiliation{\addrMilan}

\author{G. Parisi}
\affiliation{\addrRomaISC}
\affiliation{\addrRomaSAP}

\author{T. Rizzo}
\affiliation{\addrRomaISC}
\affiliation{\addrRomaSAP}

\author{J. Zinn--Justin}
\affiliation{\addrSACLAY}

\date{\today}

\begin{abstract} For a long time, the predictive limits of
perturbative quantum field theory have been limited by our inability to carry
out loop calculations to arbitrarily high order, which become increasingly
complex as the order of perturbation theory is increased.  This problem is
exacerbated by the fact that perturbation series derived from loop diagram
(Feynman diagram) calculations represent asymptotic (divergent) series which
limits the predictive power of perturbative quantum field theory.  Here, we
discuss an {\em ansatz} which could overcome these limits, based on the
observations that {\em (i)} for many phenomenologically relevant field
theories, one can derive dispersion relations which relate the large-order
growth (the asymptotic limit of ``infinite loop order'') with the imaginary part
of arbitrary correlation functions, for negative coupling (``unstable
vacuum''), and {\em (ii)} one can analyze the imaginary part for negative
coupling in terms of classical field configurations (instantons).
Unfortunately, the perturbation theory around instantons, which could lead to
much more accurate predictions for the large-order behavior of Feynman
diagrams, poses a number of technical as well as computational difficulties.
Here, we study, to further the above mentioned {\em ansatz}, correlation
functions in a one-dimensional (1D) field theory with a quartic
self-interaction and an $O(N)$ internal symmetry group, otherwise known as the
1D $N$--vector model.  Our focus is on corrections to the 
large-order growth of perturbative
coefficients, i.e., the limit of a large number of loops in the Feynman diagram
expansion. We evaluate, in momentum space, the two-loop corrections for the
two-point correlation function, and its derivative with respect to the momentum,
as well as 
the two-point correlation function with a wigglet insertion.
Also, we study the
four-point function. These quantities, computed at zero momentum transfer, enter
the renormalization-group (RG) functions (Callan--Symanzik equation) of the model.
Our calculations pave the way for further development of related methods
in field theory, and for a better understanding of field-theoretical
expansions at large order.
\end{abstract}

\pacs{11.10.Jj, 11.15.Bt, 11.25.Db, 12.38.Cy, 03.65.Db}

\maketitle

\newpage

\tableofcontents


%
%
\section{Introduction}
\label{sec1}

%
%
\subsection{Orientation}
\label{sec11}

We here lay the groundwork for the detailed analysis
of the large-order behavior of perturbation theory
for correlation functions in field-theoretical models,
pertaining to phase transitions.
Over the last decades, several steps 
have been made in the analysis 
of larger orders of perturbation theory,
for both quantum mechanical problems as well as 
field theory.
Indeed, it was Dyson who argued that,
because of vacuum instabilities induced 
for a fictitiously negative value of the 
fine-structure constant, the quantum electrodynamic
(QED) perturbation series could at best
constitute an asymptotic series~\cite{Dy1952}.
Later, this conjecture was substantiated,
and the (factorial) divergence of perturbation theory,
for both quantum mechanical oscillators~\cite{BeWu1969,BeWu1971,BeWu1973}
as well as field 
theory~\cite{BrPaZJ1977,BrLGZJ1977prd1,BrLGZJ1977prd2,BrPa1978reprinted},
was quantified both in terms of the 
power-law coefficients as well as in terms
of the additive constants in the factorial growth 
of perturbation theory at large orders.
Information regarding the leading terms in the
perturbative expansion of perturbation 
theory has been instrumental in the 
determination of critical exponents 
for the $N$-vector model, which is a $\phi^4$ 
theory with an internal $O(N)$ symmetry 
group~\cite{LGZJ1977,LGZJ1980,LGZJ1990,GuZJ1998}.

For anharmonic oscillators, 
one has been able to write down 
generalized Bohr--Sommerfeld quantization conditions
which characterize the eigenvalues, including 
instanton contributions, 
to all orders~\cite{JeZJ2004plb,ZJJe2004i,ZJJe2004ii,
JeSuZJ2009prl,JeSuZJ2010,JeZJ2011}.
From these conditions, one was able to 
infer the leading factorial divergence 
of perturbation theory,
as well as subleading corrections,
for large perturbation theory order.
Calculations were, however, restricted to 
the partition function (i.e., to the 
ground-state energy of the quantum system).

However, a decisive step which has not been 
fully clarified in the literature so far,
is the extension of the 
large-order analysis beyond leading order,
to quantities of interest other than 
the partition function. Correlation functions 
are of interest in the calculation of critical exponents.
First steps in this direction have 
been taken recently~\cite{MaPaRi2017},
with an emphasis on a scalar $\phi^4$ theory in 
two and three dimensions.
Here, we report on essential progress 
in the latter endeavor,
for a theory with an internal symmetry group 
$O(N)$, in one dimension.
First, we use a formulation of the functional 
determinant~\cite{JeZJ2011}, which allows us 
to separate the path integral around the 
non-trivial (instanton) saddle point into
integrals over the collective coordinates
(the start point of the instanton 
and the variables characterizing the internal
space of the theory), as well as
integrals over the transverse fluctuations
around the nontrivial saddle point,
in the internal symmetry group.
The functional determinant does not 
factorize into longitudinal and transverse
fluctuations (the latter being relevant to the internal space).
Second, the application of the Wick theorem
allow us to express the two-loop
corrections around the classical extremum of the action,
in terms of the longitudinal,
and transverse, propagators of the (perpendicular) 
fluctuations, where ``perpendicular'' here refers to the 
exclusion of the zero mode, which is an eigenstate
of the fluctuation operator with zero eigenvalue,
corresponding to an invariance under a collective 
coordinate. Third, the final integrations are carried out 
and lead to expressions involving 
Riemann zeta functions of even and odd integer arguments.
Eventually, we are able to carry out all integrations 
analytically. In the course of the calculations,
we find the PSLQ algorithm 
useful in the very final analytic 
steps~\cite{FeBa1992,BaPl1997,FeBaAr1999,BaBr2001}.

Here, we restrict the discussion to the 
one-dimensional case,
We put special emphasis on the partition function, 
on the two-point correlation function, 
on its derivative with respect to the momentum, 
on the two-point correlation function with a wigglet 
insertion and on the four-point correlation
function. All the correlation functions are computed 
at zero momentum transfer,
as is required for an input into the Callan--Symanzik equation.
While, in one dimension,
the field fluctuations are not strong enough 
to induce a phase transition, we clarify the 
connection of our calculations to 
the quantities entering the RG equations
in an Appendix.

This paper is organized as follows. 
We derive the functional determinant for the transformation
into collective coordinates and field fluctuations,
for the quartic $O(N)$ theory, in Sec.~\ref{sec2}.
The formalism is applied to the calculation of the 
imaginary part of the ground-state resonance energy
(i.e., to the partition function in the large-$\beta$ limit).
We use a normalization which makes the field equation 
for the instanton (classical) field configuration 
assume a particularly simple form 
[see Eq.~\eqref{SON} below].
The path integral Jacobian is derived with a 
particular emphasis on the non-factorization
of the longitudinal and transverse fluctuations.
In Sec.~\ref{sec3}, we continue with the calculation of
the two-point, and four-point, functions
as well as the derivative of the two-point function
at zero momentum transfer, and 
the wigglet insertion.
All of these functions enter the Callan--Symanzik~\cite{Ca1970,Sy1970}
renormalization-group (RG) equations.
Three appendices complement our investigations.
In Appendix~\ref{appendixb},
we supply an integral table which is useful for
the calculation of the propagator integrals.
Appendix~\ref{appendixc} is devoted to 
the connection of the correlation
functions at zero momentum, investigated here,
and the Callan--Symanzik equation.

%
%
\begin{figure}[t!]
\begin{center}
\begin{minipage}{1.0\linewidth}
\begin{center}
\includegraphics[width=0.91\textwidth]{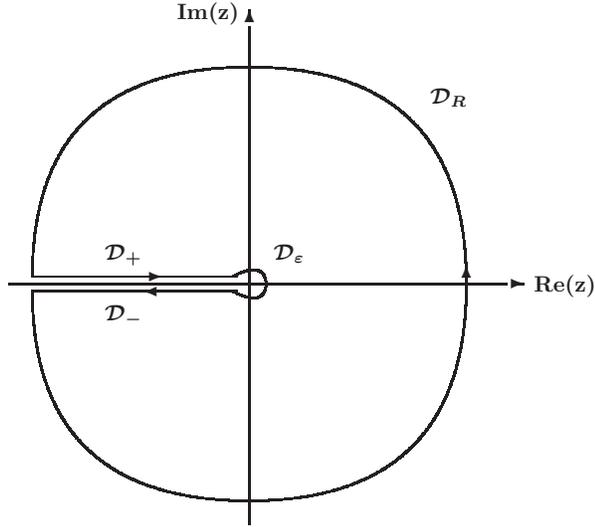}
\caption{ \label{figg1}
The complex integration path $\calD$ encircling the branch cut of the 
Green function.}
\end{center}
\end{minipage}
\end{center}
\end{figure}

%
%
\subsection{Large--Order Behavior and Analyticity}
\label{sec12}

A central point of our investigations
is the connection between the 
low-order behavior of the imaginary part of a $ n $-point
correlation function $ \sfG(g) $ and the large-order behavior of
its real part. Let us consider a generic 
Green function $\sfG(g)$ which 
is analytic in all the complex plane except on
the negative real axis. We can apply the Cauchy theorem as follows,
\begin{equation}
\sfG(g)=\frac{1}{2\pi \ii} \; \oint_D \dd z \, \frac{\sfG(z)}{z-g} \,,
\end{equation}
where $D$ is the path in the complex plane encircling the 
branch cut on the negative real axis,
and $g$ is the reference argument where the 
function $\sfG(g)$ is to be evaluated. 
The path $\calD$ can be decomposed into four 
contributions (see Fig.~\ref{figg1}),
\begin{equation}
\calD = \calD_R + \calD_\epsilon + \calD_+ + \calD_- \,. 
\end{equation}

The contributions over the paths $\calD_R$ 
and $\calD_\epsilon$ vanish identically,
and the only remaining contributions 
come from the paths $\calD_+$ and $\calD_-$.  We can then
write
\begin{align}
\label{dispersion}
\sfG(g) =& \; \frac{1}{2\pi \ii} \, \int_{-\infty}^0  \dd z \,
\frac{\textrm{disc}\,\sfG(z)}{z-g} \,,
\\[0.1133ex]
\textrm{disc} \,\sfG(z) = & \;
\lim_{\epsilon \to 0} \; 
[ \sfG(z+\ii\epsilon) - \sfG(z-\ii\epsilon) ] 
\\[0.1133ex]
=& \; -2 \, \ii \,\textrm{Im} \, \sfG(z - \ii \epsilon) \,,
\end{align}
where the discontinuity of $\sfG(z)$ on the cut is given by
$\textrm{disc}\,\sfG(z)$. In the following,
we will understand $\sfG(z)$ for $z < 0$ (on the cut)
as the value of $\sfG(z)$ obtained when $z$ acquires
an infinitesimal negative imaginary part.
Expanding the relation~\eqref{dispersion} in $z$,
we obtain
\begin{align}
\sfG(g) =& \; \sum_{K=0}^{\infty} \sfG_K \, g^K 
= -\frac{1}{\pi}\int_{-\infty}^0 \dd z \, 
\frac{\textrm{Im} \, \sfG(z - \ii \epsilon)}{z-g} 
\nonumber\\[0.1133ex]
=& \; - \frac{1}{\pi}\sum_{K=0}^{\infty}
g^K \int_{-\infty}^0 \dd z \frac{\textrm{Im} \, \sfG(z - \ii \epsilon)}{z^{K+1}} \,.
\end{align}
So, we find an integral representation for 
the perturbative coefficient of order $K$, 
of the $n$-point correlation function.
The minus sign is consistent with Eq.~(10) of Ref.~\cite{JeSuZJ2009prl}
and with Eq.~(2.31b) of Ref.~\cite{JeSuZJ2010}; note that, however,
the resonance energy in the cited publications was identified with
an infinitesimal positive imaginary part of the coupling.

We refer to the perturbative coefficient of order $K$ 
as $\sfG_K$, and write
\begin{align}
\sfG_K =& \; -\frac{1}{\pi} \int_{-\infty}^0 \dd g \,
\frac{\textrm{Im} \, \sfG(g - \ii \epsilon)}{g^{K+1}} 
\nonumber\\[0.1133ex]
=& \; \frac{(-1)^K}{\pi}\int_0^{\infty} \dd g \,
\frac{\textrm{Im} \, \sfG(-g - \ii \epsilon)}{g^{K+1}}.
\label{GkM}
\end{align}
From this equation,
we can understand the importance of knowing the value of the
imaginary part of the correlation function for small and negative value of the
coupling parameter $g$. In fact, the large order behavior of the series, i.e.
$\sfG_K$ for $K$ large, is dominated by the values of 
$\textrm{Im} \, \sfG(g)$ at
small and negative values of $g$.  

In the following, 
we will find that the 
imaginary part of a generic $n$-point function $\sfG(g)$ 
involves, in leading order, a factor $(-g)^{(-N-1+D)/2}$
from the leading-order Jacobian, given in
Eq.~\eqref{leadingJac}. We anticipate that 
a factor $(-g)^{(-N-1)/2}$
will be obtained from the $N-1$ collective coordinates
inside the $O(N)$ symmetry group, which 
give rise to the $(N-1)$th power of the classical 
field configuration in the Jacobian.
In $D$ dimensions, one has $D$ additional collective
coordinates describing translation invariance of the 
instanton in the $D$ spatial directions~\cite{BrPa1978reprinted}.
(For the current investigation, one has $D=1$.)

A further factor $(-g)^{-n/2}$ stems from the 
$n$ classical field configurations 
in the $n$-point function. 
However, additional classical field configurations
can be introduced into the 
leading-order expressions by mass derivatives,
as is evident from the discussion of the 
wigglet insertion into the two-point function
(see Sec.~\ref{sec39}). 
In general, our expressions for the 
imaginary part of a generic correlation function
$\textrm{Im} \, \sfG(g)$ have the following structure,
\begin{align}
\textrm{Im} \, \sfG(g) =& \;
c(N,D)\,(-g)^{-(n+N+D-1)/2} \exp\left( \frac{A}{g} \right) 
\nonumber\\[0.1133ex]
& \; \times 
\left[1+d(N,D)\,g+O(g^2)\right]
\label{ImGM} \,,
\qquad 
g < 0 \,,
\end{align}
where $c(N,D)$ and $d(N,D)$ are constants,
and $n$ is the number of coordinates 
entering the Green function. 
We here concentrate on the Fourier transform.
In one dimension, we find that $A = 4/3$
in our conventions of the Euclidean action~\eqref{SON}.
Inserting Eq. (\ref{ImGM}) in Eq. (\ref{GkM}), we get
\begin{equation}\begin{split}
\label{genexp}
\sfG_K =& \; c(N, D) \, \frac{(-1)^K}{\pi} \,
\int_0^{\infty} \dd g\,\frac{\ee^{-A/g}}{g^{K+(n+N+D+1)/2}} 
\\
& \; \times
\left[1-g\,d(N, D)\right] \\
=& \; \frac{c(N, D)}{\pi} \, 
 \Gamma \left(K+b\right)\left( \frac{1}{A} \right)^{(n+N+D-1)/2} \; 
\left(-\frac{1}{A}\right)^K \;
\\
& \; \times \left\{ 1 - \frac{A \, d(N, D)}{K + b - 1} \right\} \,,
\qquad
b=\frac{n + N + D - 1}{2}\,.
\end{split}\end{equation}
For large $K$, we can replace $K - 1 + b \to K$ in the 
denominator of the second term and 
identify the $1/K$-correction.
We also note the asymptotic expansion
\begin{equation}
\frac{\Gamma(K+b)}{\Gamma(K+1)} = 
K^{b-1} \, \left[ 1 + \frac{b (b-1)}{2 K} + 
\calO( 1/K^2 ) \right] \,,
\end{equation}
which can be used in order to bring the 
leading term in the expression~\eqref{genexp} into the familiar 
form $C \, K^{b-1} \, B^K \, \Gamma(K+1)$, with suitable coefficients $C$ and $B$.

For our calculations as reported below,
it is absolutely decisive to observe the
connection of the perturbative correction 
about the instanton of relative order $g$, 
given by Eq.~\eqref{ImGM},
and the subleading $1/K$ correction 
to the leading factorial growth of the 
perturbative coefficients, given in Eq.~\eqref{genexp}. 
We shall evaluate the coefficients
$d(N,D)$ by two-loop perturbation theory 
about the instanton configurations.

%
%
\section{Quartic Theory with 
\texorpdfstring{$\maybebm{O(N)}$}{O(N)} Symmetry}
\label{sec2}

%
%
\subsection{Euclidean Action}
\label{sec21}

We here follow Ref.~\cite{JeZJ2011} in the 
derivation of the $O(N)$ functional determinant,
using a field normalization which allows us to express
the field equations in a particularly simple 
analytic form.
For the $O(N)$ one-dimensional field theory, we use the action in the form,
\begin{align}
\label{SON}
{\calS}[\und q(t)] =&  \int \dd t \, \left[ 
\frac12 \, \left( \frac{\partial \und q(t)}{\partial t} \right)^2 +
\frac12 \, \und q^2(t) + \frac{g}{4} \, \und q^4(t) \right] \,,
\nonumber\\[0.1133ex]
\und q(t) =&\; \{ q_1(t), \dots, q_N(t) \} = q_\alpha(t) \, \und e_\alpha\,,
\end{align}
where an $N$-vector in the internal space is 
denoted by underlining,
and for completeness, we remark that $\und q^4(t)$
is a shorthand notation for 
$\left[ \und q^2(t) \right]^2$.
By group symmetry, for the classical field
configuration, we can pick a specific direction 
$\undu$ in the internal space, for the 
reference instanton configuration
(note, however, that an averaging over the 
possible orientations of $\undu$ is necessary at the 
end of the calculation, as discussed in the following).
The classical field configuration is found as
\begin{equation}
\label{ONclass_path}
\und q_{\rm cl}(t) = \undu \, \sqrt{-\frac{1}{g}} \, \xi_{\rm cl}(t) \,,
\qquad
\xi_{\rm cl}(t) = \frac{\sqrt{2}}{\cosh(t)} \,,
\end{equation}
which implies the existence of $N$ collective
coordinates, namely, one time translation parameterized
by $t_0$, and $N-1$ rotations in the internal 
space, leading to displacements orthogonal to the 
reference vector $\undu$.

%
%
\subsection{Fluctuation Operator}
\label{sec22}

By definition, 
the first functional derivative of the action with respect to 
$q_\beta(t)$, 
\begin{equation}
\label{I4N}
\left.
\frac{\delta {\calS}}{\delta q_\beta(t)} 
\right|_{\und q = \und q_\cl} = 
\left. \left( - \frac{\partial^2}{\partial t^2} + 
1 + g \; q_\gamma(t) \, q_\gamma(t) \right) \, q_\beta(t) 
\right|_{\und q = \und q_\cl} 
\end{equation}
vanishes at the classical path.
This resulting equation is solved by Eq.~\eqref{ONclass_path}, 
in view of the identity
\begin{equation}
\left( - \frac{\partial^2}{\partial t^2} +
1 - \xi_{\rm cl}(t)^2 \right) \, \xi_{\rm cl}(t) = 0 \,.
\end{equation}

The second functional derivative at the 
classical path gives the fluctuation operator,
for which we give a number of useful equivalent
representations,
\begin{align}
\label{defMalphabeta}
\bfM_{\alpha\beta}(t,t') =& \;
\left. \frac{\delta {\calS}}{\delta q_\beta(t) \, \delta q_\alpha(t')} 
\right|_{\und q = \und q_\cl} 
\nonumber\\[0.1133ex]
=& \; \delta(t-t') \, 
\left[ u_\alpha \; u_\beta \;
\left( - \frac{\partial^2}{\partial t^2} + 1 - 
3 \, \xi_{\rm cl}^2(t)\right)
\right.
\nonumber\\[0.1133ex]
& \; \left. + \delta_{{\TT}, \alpha \, \beta} \;
\left( - \frac{\partial^2}{\partial t^2} + 1 -
\xi_{\rm cl}^2(t)\right) 
\right] 
\nonumber\\[0.1133ex]
=& \; \delta(t-t') \, \left[ 
u_\alpha \, u_\beta \; \bfM_\LL(t) 
+ \delta_{{\TT},\alpha \beta} \; \bfM_\TT(t) \, 
\right] 
\nonumber\\[0.1133ex]
=& \; \delta(t-t') \, \bfM_{\alpha\beta}(t)\,.
\end{align}
Here, the transverse $\delta$ function is given as
\begin{equation} 
\delta_{{\TT},\alpha \beta} =
\delta_{\alpha \beta} - u_\alpha \, u_\beta \,,
\end{equation}
and we have defined 
the longitudinal (L) and transverse (T) fluctuation
operators as
\begin{subequations}
\label{M4LT}
\begin{align}
\bfM_{\LL}(t) =& \; -\frac{\partial^2}{\partial t^2} + 1 - 
\frac{6}{\cosh^2(t)} \,,
\\[0.1133ex]
\bfM_{\TT}(t) =& \;
-\frac{\partial^2}{\partial t^2} + 
1 - \frac{2}{\cosh^2(t)} \,.
\end{align}
\end{subequations}

The fluctuation operator $\bfM_{\LL} = \bfM$ 
parameterizes the longitudinal fluctuations 
(in the initially chosen direction $\und u$ of the instanton),
whereas $\bfM_{\TT}$ 
describes the transverse fluctuations
(transverse to the initially chosen direction of the instanton).
An illustrative remark is in order.
We define the domain of the operators 
$\bfM_{\LL}$ and $\bfM_{\TT}$
so that respective zero modes are excluded.
Thus, in our notation, the operators 
$\bfM_{\LL}$ and $\bfM_{\TT}$ are invertible.
In order to denote the exclusion of the zero mode,
the symbols $\bfM_{\LL}^\perp$ and $\bfM_{\TT}^\perp$
have been used in Ref.~\cite{JeZJ2011}.
Because the longitudinal fluctuation operator
fulfills $\bfM_\LL = \bfM$, where $\bfM$ is the 
fluctuation operator for the scalar theory,
we have $\bfDelta_\LL = \bfDelta$.
The inverse of $\bfM_{\alpha\beta}$
is $\bfDelta_{\alpha\beta}$, with
\begin{equation}
\label{Delta_alpha_beta}
\bfDelta_{\alpha\beta} = 
u_\alpha u_\beta \bfDelta_\LL
+ \delta_{\TT, \alpha \beta} \, \bfDelta_\TT \,.
\end{equation}
The longitudinal and 
the transverse propagators
$\bfDelta_\LL$ and
$\bfDelta_\TT$ can 
be calculated analytically~\cite{JeZJ2011},
\begin{align}
\label{Delta4xy}
\bfDelta_\LL(t_1,t_2) = & \; \frac{1}{4} \Theta(t_1-t_2) 
\frac{\sinh t_1\sinh t_2}{\cosh^2 t_1\cosh^2 t_2} \, f(t_1,t_2) 
\nonumber\\
& \; + (t_1 \leftrightarrow t_2) \,,
\\[0.1133ex]
f(t_1,t_2) =& \;
3 t_2 - 3 t_1 -1 + \ee^{t_2} \frac{ 3\sinh t_2 - 2\cosh t_2}{\tanh t_2}
\nonumber\\[0.1133ex]
& \; + \ee^{-t_1} \frac{3\sinh t_1+2\cosh t_1}{\tanh t_1} \,,
\\[0.1133ex]
\label{DeltaTxy}
\bfDelta_\TT(t_1, t_2) =& \; \frac14 \Theta(t_1-t_2) \,
g(t_1,t_2) +
(t_1 \leftrightarrow t_2) \,,
\\[0.1133ex]
g(t_1,t_2) =& \;
\left( \frac{t_2-t_1-1}{\cosh t_1 \, \cosh t_2} +
\frac{\ee^{-t_1}}{\cosh t_2} +
\frac{\ee^{t_2}}{\cosh t_1}\right)  \,.
\end{align}
A remark is in order.
Namely, a comparison of Eq.~\eqref{I4N} to~\eqref{M4LT} reveals that the 
scalar instanton configuration $\xi_{\rm cl}(t)$ constitutes a 
zero mode of the transverse part of the fluctuation operator.
The instanton path $\xi_\cl(t)$ has no zero. Therefore, when interpreted 
as a quantum mechanical wave function (eigenfunction of the fluctuation
operator), it is clear that the instanton path represents the ground state
of the transverse fluctuation operator. 
Thus, the ground state of the transverse fluctuation operator
has zero eigenvalue, which implies that all other eigenvalues
are manifestly greater than zero. The spectral determinant of the 
transverse fluctuation operator therefore is positive.

%
%
\subsection{Path Integral Jacobian}
\label{sec23}

Even though the problem of the calculation
of the functional determinant has been outlined
in Ref.~\cite{JeZJ2011},
we here revisit the derivation, with an emphasis on those 
aspects of the path integral Jacobian which are important 
for the calculation of correlation functions.
The appropriate decomposition of the path 
reads as follows,
\begin{equation}
\label{decomp_path}
{\underline q}(t) = 
{\underline u}\, [ q_{\rm cl}(t - t_0) + \chi_\LL(t - t_0) ] 
+ \undchi_\TT(t - t_0) \,,
\end{equation}
where ${\underline u}$ is a time-independent unit vector, 
${\underline u}^2=1$, chosen to point into a 
specific direction of the $(N-1)$-dimensional
unit sphere $S_{N-1}$ embedded in $N$-dimensional space. 
Furthermore, the longitudinal and transverse variations
$\chi_\LL(t)$ and $\chi_\TT(t)$ are assumed to be 
orthogonal to their respective zero modes, i.e.,
$\undu \cdot \undchi_\TT(t - t_0) = 0$.
The variable $t_0$ takes the role of a collective coordinate.
Throughout this paper, we denote vectors in the internal
symmetry space by underlining.

In order to carry out the calculation
(see Sec.~5 of Ref.~\cite{JeZJ2011}),
one has to observe that the path decomposition~\eqref{decomp_path},
under the shift $t \to t - t_0$, 
breaks both time translation and $O(N)$ invariance,
by singling out a specific direction $\und u$ in the 
internal space as well as a start time $t_0$ for the instanton.
The collective coordinates are the $N-1$ coordinates $\tau_i$
which parameterize the sphere $S_{N-1}$,
as well as the time parameter $t_0$.
One finds
\begin{multline}
\label{funcdetONprep}
\int [\dd \und q(t)] F[\und q(t)] = 
\left( \frac{1}{\sqrt{2 \pi}} \right)^N \;
\int \dd t_0 \; 
\prod_{i=1}^{N-1} \int \dd \tau_i 
\\
\times \int [\dd \chi_{\LL}(t)]  
\int [\dd \und \chi_{\TT}(t)] \, \calJ[\und q(t)] \,
\\
\times F\left[ {\underline u}\, [ q_{\rm cl}(t - t_0) + \chi_\LL(t - t_0) ] + 
\undchi_\TT(t - t_0) \right]  \,,
\end{multline}
where the Jacobian $\calJ(\und q) = \calJ[\und q(t)]$
has the representation
\begin{align}
\label{eJacobON}
\calJ(\und q) =& \;
\frac{J({\underline q})}%
{\sqrt{ J({\underline q}_{\rm cl}) }} \,,
\nonumber\\[0.1133ex]
J({\underline q}) =& \;
\det \left( \int\dd t\,
\frac{\partial {\underline q}}{\partial c_i} \cdot
\frac{\partial {\underline q}_{\rm cl}}{\partial c_j} \right)\,,
\qquad
c_i = (t_0, \tau_i) \,.
\end{align}
The collective coordinates for time translations $t_0$ 
and the collective coordinates for rotations that
parameterize $S_{N-1}$, which are denoted as $\tau_i$ 
($i=1,\ldots, N-1$), are summarized in the vector $c_i$.
It is crucial to carefully analyze 
the dependence on the collective coordinate $t_0$,
for the path as well as the Jacobian,
in the calculation of correlation functions.
Furthermore, the identification of the
path in terms of the argument $t - t_0$
(rather than $t + t_0$, 
as in Sec.~5 of Ref.~\cite{JeZJ2011}) 
serves to illustrate the role of 
$t_0$ as the ``reference start point'' of the 
classical path.

The rationale behind the transformation~\eqref{funcdetONprep} is as follows.
We start from the path integral over closed paths $\oint [\dd \und q(t)]$.
There are $N-1$ collective coordinates in the internal space of the $O(N)$
theory, and one collective coordinate describing the time translation of the
longitudinal instanton. This means that 
there are $N$ collective coordinates in total; the
exclusion of these from the remaining path integral leads to a factor $(2
\pi)^{-N/2}$. In the 
remaining integral over the fluctuations $\oint [\dd \chi_\LL(t)]$,
the longitudinal zero mode corresponding to the instanton path is excluded,
leading to convergent expressions for the Gaussian path integral expectation
values.  The same applies to the integration over the transverse fluctuations
$\oint [\dd \und \chi_\TT(t)]$, where we exclude the transverse zero
mode, to be discussed below,
in all directions perpendicular to the fixed vector $\und u$ in the
internal space. 

The path ${\underline q}(t)$ is the sum of the classical 
path ${\underline u}\,q_{\rm cl}(t - t_0)$ and 
two sums over longitudinal fluctuations ($\LL$), 
and transverse fluctuations ($\TT$). The transverse fluctuations
may point in any of the $N-1$ available directions.
The $N-1$ vectors $\und e_2, \dots, \und e_N$ parameterize the 
transverse fluctuations, orthogonal to $\undu$
(where $\und u$ can point into any direction in 
the internal space).
We also set
\begin{equation}
{\underline q}(t)=
{\und q}_{\LL}(t - t_0) +
{\und q}_{\TT}(t - t_0) \,,
\end{equation}
with self-explanatory definitions for the longitudinal
component ${\und q}_{\LL}(t - t_0) = 
{\underline u} \, q_{\LL}(t - t_0)$ and 
the transverse component ${\und q}_{\TT}(t - t_0)$.
The function $\dot q_{\rm cl}(t)$ is the zero mode 
of the longitudinal fluctuation operator, whereas 
$q_{\rm cl}(t)$ is the zero mode
of the transverse fluctuation operator.
The conditions that the zero-modes should be omitted therefore 
read 
\begin{equation}
\label{eqTortho}
\int\dd t\,\dot q_{\rm cl}(t)\bigl( q_{\LL}(t)-  q_{\rm cl}(t)\bigr) = 0\,,
\quad
\int\dd t\,q_{\rm cl}(t) {\underline q}_{\TT}(t)= \underline{0}\,,
\end{equation}
where the first condition comes from translations and the second from $O(N)$
rotations. The next step is to 
calculate the matrix elements relevant for the 
expression~\eqref{eJacobON},
\begin{subequations}
\begin{align}
\label{Jcalc}
J(\vec q) = &\;
\det \left( \begin{array}{cc} \calA & \calB^\rmT \\ \calC & \calD
\end{array} \right) 
\nonumber\\[0.1133ex]
=& \; 
\left(\calA - \calB^\rmT \, \calD^{-1} \, \calC \right) \,
\det(\calD) \,,
\\[0.1133ex]
\calA =& \; \int\dd t\,\frac{\partial \und q(t - t_0)}{\partial t_0} \cdot 
\frac{\partial \und q_{\rm cl}(t - t_0)}{\partial t_0} 
\nonumber\\[0.1133ex]
=& \; \int\dd t\,\dot q_{\LL}(t) \, \dot q_{\rm cl}(t) \,,
\\[0.1133ex]
\calB^\rmT_j =& \; \int\dd t \,
\frac{\partial {\underline q}(t - t_0)}{\partial t_0} \cdot
\frac{\partial {\underline q}_{\rm cl}(t - t_0)}{\partial \tau_j} 
\nonumber\\[0.1133ex]
=& \; - \frac{\partial {\underline u}}{\partial \tau_j} \cdot
\int\dd t\,\dot {\underline q}_{\TT}(t) \; q_{\rm cl}(t) \,,
\\[0.1133ex]
\calC_i =& \;
\int\dd t\,
\frac{\partial {\underline q}_{\rm cl}(t - t_0)}{\partial t_0} \cdot
\frac{\partial {\underline q}(t - t_0)}{\partial \tau_i} 
\nonumber\\[0.1133ex]
=& \; -{\underline u}\cdot\int\dd t\, 
\dot q_{\rm cl}(t) \, 
\frac{\partial {\underline q}_{\TT}(t)}{\partial \tau_i} \,,
\\[0.1133ex]
\calD_{ij} =& \; \int\dd t \, 
\frac{\partial {\underline q}(t - t_0)}{\partial \tau_i} \cdot
\frac{\partial {\underline q}_{\rm cl}(t - t_0)}{\partial \tau_j} 
\nonumber\\[0.1133ex]
=& \; g_{ij} \, \int\dd t\,q_{\LL}(t) \, q_{\rm cl}(t) \,.
\end{align}
\end{subequations}
Here, $\calB^\rmT$ is a row vector, 
$\calC$ is a column vector,
$\calA$ is a number,
while $\calD$ is an $(N-1) \times (N-1)$ matrix.
We have introduced the metric 
\begin{equation}
g_{ij} = \frac{\partial {\underline u}}{\partial \tau_i} \cdot
\frac{\partial {\underline u}}{\partial \tau_j} \,
\end{equation}
on the sphere $S_{N-1}$.
We can write in the leading order,
\begin{align}
\label{leadingJac}
\calJ[\undq(t)] \approx & \; \sqrt{J[\undq_{\rm cl}(t)]} =
\sqrt{ \det\left( g_{ij} \right) } \,
|| {\underline q}_{\rm cl} ||^{N-1} \,
|| \dot {\underline q}_{\rm cl} || 
\nonumber\\[0.1133ex]
=& \; \sqrt{ \det\left( g_{ij} \right) } \,
\left( - \frac{3 A}{g} \right)^{(N-1)/2} \,
\left( - \frac{A}{g} \right)^{1/2} \,,
\end{align}
where $|| f ||$ is the norm $[\int_{-\infty}^\infty {\rm d} t \, f(t)^2]^{1/2}$.
A very useful representation is obtained upon division
by the square root of the determinant of the metric
in the internal space, which 
in view of Eq.~\eqref{Jcalc} is contained
in the term $\det (\calD)$. One finds
\begin{subequations}
\label{JACRES}
\begin{align}
\frac{\calJ[\und q(t)]}{\left(\det g_{ij}\right)^{1/2}} 
=& \;
\left( \frac{J[\und q_\cl(t)]}{\det g_{ij}} \right)^{1/2} \,
\frac{J[\und q(t)]}{J[\und q_\cl(t)]} 
\\[0.1133ex]
=& \; \left( - \frac{3 A}{g} \right)^{(N-1)/2} \,
\left( - \frac{A}{g} \right)^{1/2} \,
\frac{J[\und q(t)]}{J[\und q_\cl(t)]} \,,
\nonumber\\[0.1133ex]
\label{FFcorr}
\frac{J[\und q(t)]}{J[\und q_\cl(t)]}  =& \;
\frac{\left( \int\dd t\,{\underline q}(t)\cdot 
{\underline q}_{\rm cl}(t) \right)^{N-2}}%
{|| q_{\rm cl}(t) ||^{2(N-2)}} 
\frac{K}{ || \dot q_{\rm cl}(t) ||^2 \, || q_{\rm cl}(t) ||^2} \,,
\nonumber\\[0.1133ex]
K =& \; \int\dd t\,\dd t'
\left[\dot{\underline q}(t)  \cdot \dot{\underline q}_{\rm cl}(t) \,
{\underline q}(t') \cdot {\underline q}_{\rm cl}(t') 
\right. 
\nonumber\\[0.1133ex]
& \; \left. -
\dot q_{\rm cl}(t) \; \dot q_{\rm cl}(t') \;
{\underline q}_{\TT}(t) \cdot {\underline q}_{\TT}(t')\right]  \,.
\end{align}
\end{subequations}
With the help of Eq.~\eqref{funcdetONprep},
we are now in the position to write the following identity,
\begin{subequations}
\label{FFF}
\begin{align}
\label{FF1}
\int [\dd \und q(t)] F[\und q(t)] = & \;
\left( \frac{1}{\sqrt{2 \pi}} \right)^N \, \sigma_N \,
\int \dd t_0 \int [\dd \chi_{\LL}(t)] \, 
\nonumber\\[0.1133ex]
& \; \hspace{-2cm} \times 
\int [\dd \und \chi_{\TT}(t)] \,
\left< 
\left( \frac{\calJ[\und Q(t)]}{\left(\det g_{ij}\right)^{1/2}} \right) 
F\left(\und Q(t) \right) 
\right>_{S_{N-1}} \!\!\! \,,
\\[0.1133ex]
\und Q(t) =& \;
u \, \left( q_{\rm cl}(t-t_0) +
\chi_{\LL}(t-t_0) \right) 
\nonumber\\[0.1133ex]
& \; + \und \chi_{\TT}(t - t_0) \,,
\\[0.1133ex]
\label{FF2}
\sigma_N =& \; 
\prod_{i=1}^{N-1} \int \dd \tau_i \, \left(\det g_{ij}\right)^{1/2} =
\frac{2 \; \pi^{N/2}}{\Gamma(N/2)} \,,
\\[0.1133ex]
\label{FF3}
\langle f(\undu) \rangle_{S_{N-1}} =& \; 
\frac{1}{\sigma_N} \,
\prod_{i=1}^{N-1} \int \dd \tau_i \, \left(\det g_{ij}\right)^{1/2} 
f(\undu) \,.
\end{align}
\end{subequations}
Here, $\sigma_N$ the surface of $S_{N-1}$,
and the expression $\langle f(\undu) \rangle_{S_{N-1}}$ 
indicates the averaging of 
the test function $f(\undu)$ over the $S_{N-1}$ sphere.
As an example for the averaging process,
we indicate the formula
$\langle u_\alpha \, u_\beta \rangle_{S_{N-1}} = 
\delta_{\alpha\beta}/N$.
One important observation is that the path $\calJ[\und q(t)]$
in the Jacobian can be taken with a start time $t_0 = 0$
of the path.
This is because all integrals contributing to the Jacobian
are independent of $t_0$.
However, the decomposition~\eqref{decomp_path} is still valid; 
the path $\und q(t)$ depends on the start time $t_0$, 
and this dependence has to be figured into the integrand.

%
%
\subsection{\texorpdfstring{$\maybebm{O(N)}$}{O(N)} Quartic Oscillator}
\label{sec24}

Let us briefly review the calculation of 
the perturbative expansion of the ground-state 
energy for the $O(N)$ case, from the path integral 
representation. We write the Euclidean 
action as
\begin{align}
\label{S4_ON}
\calS[\und q(t)] = & \;
\int_{-\beta/2}^{\beta/2} \dd t_1 \!
\int_{-\beta/2}^{\beta/2} \dd t_2 \; \undq(t_1) \, 
\bfM_0(t_1,t_2) \, \undq(t_2) 
\nonumber\\[0.1133ex]
& \; + \frac14 \, g 
\int_{-\beta/2}^{\beta/2} \dd t\; \und q(t)^4 \,,
\end{align}
where the free fluctuation operator $\bfM_0$ and its
inverse $\bfDelta_0$ are given by
\begin{subequations}
\begin{align}
\label{M0}
{\bfM}_0(t_1,t_2) =& \; \delta(t_1 - t_2) \, {\bfM}_0(t_2) \,,
\qquad
{\bfM}_0(t) = - \frac{\partial^2}{\partial t^2} + 1 \,,
\\[0.1133ex]
\label{Delta0}
{\bf\bfDelta}_0 \cdot \bfM_0 =& \; \mathbbm{1} \,,
\qquad
{\bf\bfDelta}_0(t_1, t_2) = 
\frac12 \, \exp(- | t_1 - t_2 | ) \,.
\end{align}
\end{subequations}
One writes
\begin{align}
E_0(g) =& \; \lim_{\beta \to \infty} \left( -\frac{1}{\beta} \, 
\ln\left( \frac{ \calZ_0(\beta) }{ \left. \calZ_0(\beta) \right|_0 } \right) \right) + 
\frac{N}{2} \,,
\end{align}
where $\calZ_0(\beta)$ is the saddle-point expansion
of the partition function $\calZ(\beta)$,
redefined for the $O(N)$ oscillator, about the Gaussian saddle point,
and $\left. \calZ_0(\beta) \right|_0$ is obtained from $\calZ_0(\beta)$ by 
setting $g = 0$. 
The partition function can be written as follows,
\begin{align}
\left. \calZ_0(\beta) \right|_0 =
\oint [{\dd} \undq(t)] 
\exp \left[ -\tfrac12 \int \dd t \! \int \dd t' \undq(t) \bfM_0(t, t') \undq(t') 
\right]  \,,
\end{align}
where
\begin{equation}
\label{defoint}
\oint [\dd q(t)] \equiv \int_{-\infty}^\infty \dd q_0 \,
\int_{q(-\beta/2) = q_0}^{q(\beta/2) = q_0}  
\end{equation}
is a path integral over all periodic paths.
We define a normalization factor
\begin{align}
\calN =& \;
\int [{\dd} \undq(t)] \, \exp \left[ -
\frac12 \int \dd t \, \int \dd t' \, \undq(t) \, \bfM_0(t, t') \, \undq(t') 
\right] 
\nonumber\\[0.1133ex]
=& \; \frac{1}{( \det \bfM_0 )^{N/2} } \,.
\end{align}
A perturbative expansion up to the order $g^2$ leads to the result,
\begin{multline}
\frac{\calZ_0(g)}{ \left. \calZ_0(\beta) \right|_0 } = 1
- \frac{g}{4 \calN} \, \intbeta {\dd}t 
\int [{\dd} \undq(t)] \, [\undq(t)]^4 \, 
\calE[\undq(t)]
\\[2ex]
+ \frac{g^2}{32 \calN}
\intbeta\dd t \intbeta\dd t' \int [{\dd}q(t)] \, 
\undq(t)^4 \, \undq(t')^4 \, \calE[\undq(t)] \,,
\end{multline}
where
\begin{equation}
\calE[\undq(t)] = \exp \left( - \frac12 \, \int \dd t_1 \int \dd t_2 \, 
\undq(t_1) \, \bfM_0(t_1, t_2) \, \undq(t_2)  \right).
\end{equation}
We define the path integral expectation value 
$\langle Y \rangle_0$ as
\begin{equation}
\label{WEIGHT0_ON}
\left< Y \right>_0 = \left( \det \bfM_0 \right)^{N/2} \;
\int [{\dd} \und q(t)] \;  Y \; \calE[\undq(t)] \,.
\end{equation}
Application of the Wick theorem leads to
\begin{equation}
\left< \und q^4(t) \right>_0 =
N (N+2) \bigl[ \bfDelta_0(0) \bigr]^2 \, 
= \tfrac14 \, N \, (N + 2) \,,
\end{equation}
while the generalization to 
$\left< \und q^4(t) \, \undq(t')^4 \right>_0$
is straightforward. Finally, one obtains
\begin{align}
\ln \left( \frac{\calZ_0(\beta)}{ \left. \calZ_0(\beta) \right|_0 } \right) = & \; 
- \frac{\beta g}{16} \, N (N+2) 
\\[0.1133ex]
& \; + \frac{\beta g^2}{128} \, N (N+2) (2N + 5) \,,
\nonumber\\[0.1133ex]
E_0(g) = & \; \frac{N}{2} + \frac{g}{16} \, N \, (N+2)
\\[0.1133ex]
& \; - \frac{g^2}{128} \, N \, (N+2) \, (2 N + 5) \,,
\nonumber
\end{align}
where we ignore terms of order $g^3$ and higher and 
confirm the cancelation of $\beta$ in the 
expression for $E_0(g)$.

%
%
\subsection{Decay Width and Instanton}
\label{instON}

We are now in the position to present the analogous derivation of the 
leading-order result for the imaginary part of the 
ground-state resonance.
The action~\eqref{SON}, expressed in terms 
of the classical action plus fluctuations about the 
instanton configuration, becomes
\begin{multline}
\label{SONpert}
{\calS}[\und \chi(t)] = -\frac{4}{3 g} +
\frac{1}{2} \int \dd t_1 \! \int \dd t_2 
\chi_\alpha(t_1) {\bfM}_{\alpha\beta}(t_1,t_2) \chi_\beta(t_2) 
\\[0.1133ex]
- \sqrt{-g} \int \dd t 
\und \xi_{\rm cl}(t) \cdot \und\chi(t) \;\; \und \chi^2(t) 
+ \frac{g}{4} \int \dd t \; \und \chi^4(t) \,.
\end{multline}
For the calculation of the leading-order term 
in the decay width, we need the second term
on the right-hand side,
which is the term involving the fluctuation operator.
We use Eqs.~\eqref{funcdetONprep}, as well as 
Eqs.~\eqref{FF1},~\eqref{FF2} and~\eqref{FF3}.
Observe that, for the partition function,
we can simply integrate out the collective coordinate
$\int \dd t_0 = \beta$.
The leading contribution to the 
imaginary part ${\rm Im} \, E_0(g)$ for the ground-state energy of the 
$O(N)$ quartic oscillator is obtained as
\begin{align}
\label{LeadQuartONInst}
{\rm Im} \, E_0(g) \approx & \;
\lim_{\beta \to \infty} 
\left( -\frac{1}{\beta} \;
\frac{{\rm Im} \; \calZ_1(\beta)}{\calZ_{0}(\beta)} \right)
\nonumber\\[0.1133ex]
= & \; - \frac{1}{ \Gamma(N/2)} 
\frac{1}{2^{N/2}} 
\left( - \frac{ 4 }{ g } \right)^{N/2} 
\frac{ 1 }{ \sqrt{3} } 
\exp \left( \frac{4}{3 g} \right)
\nonumber\\[0.1133ex]
& \; \times
\left[ -\det\left( \frac{1}{\bfM_0} \bfM_\LL \right) \right]^{-1/2} \;
\nonumber\\[0.1133ex]
& \; \times
\left[ \det\left( \frac{1}{\bfM_0} \bfM_\TT \right) \right]^{ -\frac12(N-1) } 
\nonumber\\[0.1133ex]
= & \; - \frac{1}{ \Gamma(N/2)} \;
\left( - \frac{ 8 }{ g } \right)^{N/2} \,
\exp \left( \frac{4}{3 g} \right) \,.
\end{align}
We resolve the ambiguity in taking the square
root so that the imaginary part of the energy comes
out as negative.
The derivation in Eq.~\eqref{LeadQuartONInst}
implicitly supposes that $g$ is negative.
As we saw in Sec.~\ref{sec12},
the particular sign of the imaginary part 
chosen in Eq.~\eqref{LeadQuartONInst} corresponds to 
values of $g$ with an infinitesimal negative imaginary part.
In the derivation, we have used the results~\cite{JeZJ2011}
\begin{equation}
\label{MLM0T}
\det \left( \frac{1}{\bfM_0} \bfM_{\LL} \right) = -\frac{1}{12}\,,
\quad
\det \left( \frac{1}{\bfM_0} \bfM_{\TT} \right) = \frac{1}{4}\,.
\end{equation}

%
%
\subsection{Corrected
\texorpdfstring{$\maybebm{O(N)}$}{O(N)} Decay Width}

\begin{figure}[t!]
\begin{center}
\begin{minipage}{0.91\linewidth}
\begin{center}
\includegraphics[width=0.95\linewidth]{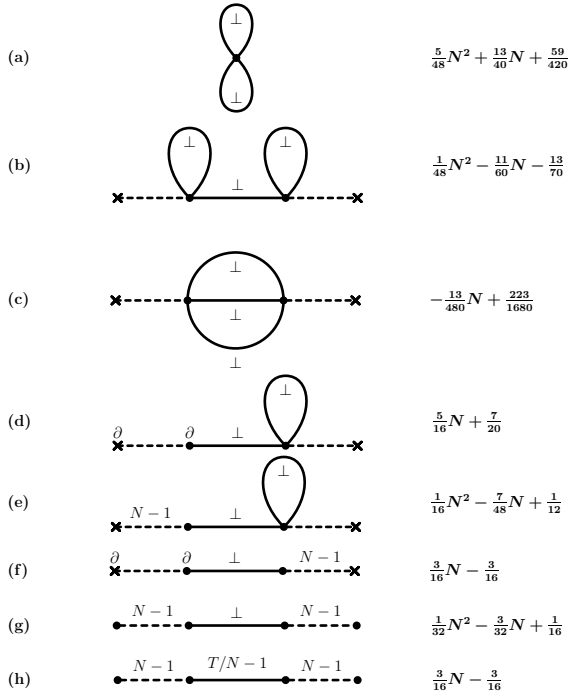}
\caption{\label{figg2} Diagrammatic representation
of several terms in a $\phi^4$ theory
with an $O(N)$ internal symmetry,
contributing to the partition function
in the infinite-$\beta$ limit and thus,
to the ground-state energy.
The contribution of the diagrams is written beside each 
contribution. The total result of order $g$ is 
of the form given in Eq.~\eqref{Z1corr}.}
\end{center}
\end{minipage}
\end{center}
\end{figure}

The key to the calculation of the corrections to the 
partition function, and (later on) to the corrections to 
the correlation functions, lies in the inclusion
of corrections from three sources:
{\em (i)} perturbative corrections from the expansion of the 
action~\eqref{SONpert}, which enters the 
exponential $\exp(-\calS[ \undchi(t)])$,
{\em (ii)} perturbative corrections from the 
expansion of the Jacobian $J[\undq(t)]/J[\undq_\cl(t)]$,
and 
{\em (iii)} perturbative corrections from the 
denominator $\calZ_0(\beta)$, in the expression
$(-1/\beta) \, {\rm Im}[\calZ_1(\beta)/\calZ_{0}(\beta)]$,
in the limit of large $\beta$.
An expansion of the exponential
$\exp\left( -{\calS}[\und \chi(t)] \right)$,
according to Eq.~\eqref{SONpert}, leads to the
correction factor $F_1$,
\begin{align}
\label{factorF1}
F_1 =& \; 
\frac{\ee^{ -{\calS}[\und \chi(t)] }}
{\left. \ee^{ -{\calS}[\und \chi(t)] } \right|_0}
= 1 +
\sqrt{-g} \, \int \dd t \und \xi_{\rm cl}(t) \cdot 
\und\chi(t) \;\; \und \chi^2(t)
\nonumber\\[2ex]
& \; 
- \frac{g}{4} \, \int \dd t\, \und \chi^4(t) 
- \frac{g}{2} \,
\left( \int \dd t \; \und \xi_{\rm cl}(t) \cdot \und\chi(t) \;\; \und \chi^2(t) \right)
\nonumber\\[2ex]
& \; \times
\left( \int \dd t' \; \und \xi_{\rm cl}(t') \cdot \und\chi(t') \;\; 
\und \chi^2(t')\right) \,,
\end{align}
where $\left. {\calS}[\und \chi(t)] \right|_0 =
\int \dd t_1 \int \dd t_2 \, 
\chi_\alpha(t_1) \, \bfM_{\alpha\beta}(t_1, t_2) \, \chi_\beta(t_2)$,
which has to be inserted inside the path integral.
The second factor $F_2$ is from the $O(N)$ Jacobian,
\begin{align}
\label{factorF2}
F_2 =& \; \frac{J[\und q(t)]}{ J[\und q_{\rm cl}(t)] } =
1 + \frac34 \sqrt{-g} \; 
\int \dd t \, \dot{\und \chi}(t) \cdot \dot{\und \xi}_{\rm cl}(t)
\nonumber\\[2ex]
& \; \hspace{-0.8cm}
+ (N-1) \, \frac{\sqrt{-g}}{4} \;
\int\dd t\,{\underline \chi}(t)\cdot {\underline \xi}_{\rm cl}(t) 
\nonumber\\[2ex]
& \; \hspace{-0.8cm} 
- \frac{3}{16} (N-1) \, g \;
\int \dd t \, \int \dd t' \,
\dot{\und \chi}(t) \cdot \dot{\und \xi}_{\rm cl}(t) 
{\und \chi}(t') \cdot {\und \xi}_{\rm cl}(t') 
\nonumber\\[2ex]
& \; \hspace{-0.8cm} 
- \frac{g}{32} (N-1) (N-2) 
\int\dd t \int\dd t'
{\underline \chi}(t)\cdot {\underline \xi}_{\rm cl}(t)
{\underline \chi}(t')\cdot {\underline \xi}_{\rm cl}(t')
\nonumber\\[2ex]
& \; \hspace{-0.8cm} + \frac{3}{16} g\; \int \dd t \, \int \dd t' \,
\dot \xi_{\rm cl}(t) \;
{\und \chi}_{\TT}(t) \cdot {\und \chi}_{\TT}(t') \;
\dot \xi_{\rm cl}(t') \,.
\end{align}
Furthermore, there is a factor from the perturbative
expansion of the denominator, which originates from the 
Gaussian saddle point (see Sec.~\ref{sec24}),
\begin{align}
\label{factorF3}
F_3 =& \; \frac{\left. \calZ_0(\beta) \right|_0}{\calZ_0(\beta)} 
= 1 + \frac{g}{4} \int \dd t\, N(N+2) \, [\bfDelta_0(0)]^2 
\nonumber\\[0.1133ex]
=& \; 1 + \frac{g}{4} \int \dd t\, 
[ 3 + 2(N-1) + (N^2-1) ] \, [\bfDelta_0(t,t)]^2 \,.
\end{align}
The latter form is very handy when it 
comes to subtracting infinities.
The final result can be written as
\begin{align}
\label{ff1}
{\rm Im} \, E_0(g) \approx & \;
- \frac{1}{ \Gamma(N/2)} \;
\left( - \frac{ 8 }{ g } \right)^{N/2} \,
\exp \left( \frac{4}{3 g} \right) 
\nonumber\\[0.1133ex]
& \; \times 
\left( 1 + {\mathcal A} + {\mathcal B} + {\mathcal C} \right) \,,
\end{align}
where the terms ${\mathcal A}$, ${\mathcal B}$, and 
${\mathcal C}$ are of order $g$,
given by
\begin{equation}
\calA = \sum_{i=1}^3 \calA_i \,,
\quad
\calB = \sum_{j=1}^2 \calB_j \,,
\quad
\calC = \sum_{k=1}^3 \calC_k \,,
\end{equation}
as defined in the following.
(These are of course different from the 
submatrices $\calA$, $\calB$ and $\calC$ used in Sec.~\ref{sec23};
we redefine the symbols $\calA$, $\calB$ and $\calC$ accordingly.)
We distinguish the terms into ${\mathcal A}$, ${\mathcal B}$, and 
${\mathcal C}$ as follows.
The ${\mathcal A}$ originate from the effective action,
i.e., from $F_1$, while infinities are removed by $F_3$.
They correspond to the first three diagrams in Fig.~\ref{figg2}.
The ${\mathcal B}$ terms contain the mixed contributions from 
the product $F_1 \times F_2$, expanded to order $(\sqrt{-g})^2 = -g$
(see the forth and fifth diagrams in Fig.~\ref{figg2}).
Terms of order $g$ in $F_2$ 
give rise to ${\mathcal C}$ (Jacobian terms,
see the sixth, seventh and 
eighth diagrams in Fig.~\ref{figg2}).

We start with the ${\mathcal A}$ term,
\begin{align}
\label{AAAterm}
{\mathcal A} = & \;
- \frac{g}{4} \, \int \dd t\, 
\left( \left< \und \chi^4(t) \right> - 
\left< \und \chi^4(t) \right>_0 \right)  
- \frac{g}{2} \, \int \dd t \int \dd t' 
\nonumber\\[0.1133ex]
& \; \times \left<  \und \xi_{\rm cl}(t) \cdot \und\chi(t) \;\; 
\und {\chi}^2(t) 
\; \und \xi_{\rm cl}(t') \cdot \und\chi(t') \;\; \und \chi^2(t') \right> \,.
\end{align}
We define $\langle \cdot \rangle$ for the $O(N)$ theory as 
\begin{align}
\label{WEIGHTLT}
& \left< X \right> = 
\left( \det \bfM_{\LL} \right)^{1/2}
\left( \det \bfM_{\TT} \right)^{N/2}
\oint [{\dd} \chi_{\LL}(t)] 
\oint [{\dd} \und \chi_{\TT}(t)] \, X
\nonumber\\[0.1133ex]
& \quad \times 
\exp \left( - \frac12 \, \int \dd t_1 \, \int \dd t_2 \;
\chi_\alpha(t_1) \, {\bfM}_{\alpha\beta}(t_1, t_2) \, 
\chi_\beta(t_2)  \right) \,.
\end{align}
The term ${\mathcal A}_1$ term is easy,
\begin{align}
{\mathcal A}_1 =& \; - \frac{g}{4} \, \int \dd t\,
\left( \left< \chi_{\LL}^4(t) \right> +
2 \left< \chi_{\LL}^2(t) \, \und \chi_{\TT}^2(t) \right>
+ \left< \und \chi_{\TT}^4(t) \right> \right) 
\nonumber\\[0.1133ex]
& \; 
+ \frac{g}{4} \, \int \dd t\, \left< \und \chi^4(t) \right>_0 \,.
\end{align}
Applying the Wick theorem, we obtain the result
\begin{align}
{\mathcal A}_1 = & \; - \frac{g}{4} \int \dd t
\left( 
3 [ \bfDelta_{\LL}^2 (t,t) 
- \bfDelta_0^2(t,t) ] 
\right.
\nonumber\\[0.1133ex]
& \; 
+ 2 (N-1) [\bfDelta_{\LL}(t,t) \, \bfDelta_{\TT}(t,t) 
- \bfDelta_0^2(t,t) ] 
\nonumber\\[0.1133ex]
& \; 
\left. 
+ (N^2 -1) [ \bfDelta_{\TT}^2 (t,t) - \bfDelta_0^2(t,t) ] 
\right) 
\nonumber\\[0.1133ex]
=& \; g \, 
\left( \frac{5}{48} N^2 + \frac{13}{40} \, N + \frac{59}{420} \right) \,.
\end{align} 
For the second term in Eq.~\eqref{AAAterm}, one has
\begin{equation}
\left<  \und \xi_{\rm cl}(t) \cdot \und\chi(t) \;\; \und \chi^2(t)
\; \und \xi_{\rm cl}(t') \cdot \und\chi(t') \;\; \und \chi^2(t') \right> =
T_1 + T_2 \,,
\end{equation}
where
\begin{subequations}
\begin{align}
T_1 = & \; \xi_{\rm cl}(t) \;
\left[ 3 \bfDelta_{\LL}(t,t) + (N-1) \, \bfDelta_{\TT}(t,t) \right] \,
\bfDelta_{\LL}(t,t') 
\nonumber\\[2ex]
& \; \times 
\left[ 3 \bfDelta_{\LL}(t',t') + (N-1) \, \bfDelta_{\TT}(t',t') \right] \,
\; \xi_{\rm cl}(t') \,,
\\[2ex]
T_2 =& \; 2 \, \xi_{\rm cl}(t) \; \bfDelta_{\LL}(t,t') \;
\left( 3 \bfDelta^2_{\LL}(t,t') \right.
\nonumber\\[2ex]
& \; \left. + (N-1) \; \bfDelta^2_{\TT}(t,t') \right) \,
\xi_{\rm cl}(t') \,.
\end{align}
\end{subequations}
The expression $T_1$ generates the term $\calA_2$
[see the diagram in Fig.~\ref{figg2}(b)],
while the expression $T_2$ generates the term $\calA_3$
[see the diagram in Fig.~\ref{figg2}(c)],
\begin{align}
{\mathcal A}_2 = & \; 
g \, \left( \frac{1}{48} N^2 - \frac{11}{60} \, N - \frac{13}{70} \right) \,,
\\[0.1133ex]
{\mathcal A}_3 = & \; 
g \, \left( - \frac{13}{480} \, N + \frac{223}{1680} \right) \,.
\end{align}
We now turn our attention to the 
$\calB$ terms, which are generated by mixed contributions 
from $F_1$ and $F_2$. In fact,
there are two terms in $F_2$ proportional to $\sqrt{-g}$,
one of them being proportional to $(N-1)$. 
When multiplied by the term of order $\sqrt{-g}$ from $F_1$, 
these generate two mixed Feynman diagrams. 
The corresponding expression for the diagram in Fig.~\ref{figg2}(d) reads
\begin{align}
{\mathcal B}_1 
=& \; \frac{3g}{4} \int \dd t \int \dd t' 
\left< \dot{\und \chi}(t) \cdot \dot{\und \xi}_{\rm cl}(t)
\und \xi_{\rm cl}(t') \cdot \und\chi(t') \;\; \und \chi^2(t')
\right> 
\nonumber\\[2ex]
=& \; g \left( \frac{5}{16} \, N + \frac{7}{20} \right) \,.
\end{align}
Furthermore, we have the expression for the 
diagram in Fig.~\ref{figg2}(e), 
\begin{align}
{\mathcal B}_2 =& \;
(N-1) \, \frac{g}{4} \; \int\dd t\, \int \dd t'\, 
\nonumber\\[0.1133ex]
& \; \times \left< {\underline \chi}(t)\cdot {\underline \xi}_{\rm cl}(t)
\und \xi_{\rm cl}(t') \cdot \und\chi(t) \;\; \und \chi^2(t') \right> 
\\[2ex]
= & \; g \, \left(
\frac{1}{16} \, N^2 - \frac{7}{48} \, N + \frac{1}{12} \right) \,.
\nonumber
\end{align}
There are three more terms generated by the terms of order $g$ in the 
$O(N)$ Jacobian. The first of these is given in Fig.~\ref{figg2}(f) and 
reads
\begin{align}
{\mathcal C}_1 =& \; -\frac{3(N-1)g}{16}
\int \dd t \int \dd t' 
\left< \dot{\und \chi}(t) \cdot \dot{\und \xi}_{\rm cl}(t) 
{\und \chi}(t') \cdot {\und \xi}_{\rm cl}(t') \right>
\nonumber\\[2ex]
=& \; g \, \left( \frac{3}{16} \, N - \frac{3}{16} \right) \,.
\end{align}
The diagram given in Fig.~\ref{figg2}(g) gives rise to
\begin{align}
{\mathcal C}_2 =& \;
- \frac{g}{32} \, (N-1) \, (N-2) \,
\int\dd t \, \int\dd t' \, 
\nonumber\\[0.1133ex]
& \; \times \left< 
{\underline \chi}(t)\cdot {\underline \xi}_{\rm cl}(t)
{\underline \chi}(t')\cdot {\underline \xi}_{\rm cl}(t')
\right>
\\[2ex]
=& \; g \, \left(
\frac{1}{32} \, N^2 - \frac{3}{32} \, N + \frac{1}{16} 
\right) \,.
\nonumber
\end{align}
To complete the list, we analyze the diagram in Fig.~\ref{figg2}(h),
\begin{align}
{\mathcal C}_3 =& \;
\frac{3 g}{16} 
\int \dd t \, \int \dd t' \,
\left<
\dot \xi_{\rm cl}(t) \;
{\und \chi}_{\TT}(t) \cdot {\und \chi}_{\TT}(t') \;
\dot \xi_{\rm cl}(t') 
\right>
\nonumber\\[2ex]
=& \; g \, \left( \frac{3}{16} \, N - \frac{3}{16} \right) \,.
\nonumber
\end{align}
The result for the imaginary part of the $O(N)$ ground state resonance 
finally is obtained as 
\begin{align}
\label{Z1corr}
{\rm Im} \, E_0(g) =& \;
- \frac{1}{ \Gamma(N/2)} \;
\left( - \frac{ 8 }{ g } \right)^{N/2} \,
\exp \left( \frac{4}{3 g} \right) 
\nonumber\\[2ex]
& \; \times \left( 1 
+ \sum_{i=1}^3 {\mathcal A}_i 
+ \sum_{j=1}^2 {\mathcal B}_j 
+ \sum_{k=1}^3 {\mathcal C}_k 
\right) 
\nonumber\\[2ex]
=& \;
- \frac{1}{ \Gamma(N/2)} \;
\left( - \frac{ 8 }{ g } \right)^{N/2} \,
\exp \left( \frac{4}{3 g} \right) \,
\nonumber\\[2ex]
& \; \times \left[ 1 + 
g \, \left( \frac{7}{32} N^2 + \frac{9}{16} N + \frac{5}{24} \right)
\right] \,.
\end{align}
This result is relevant for $g<0$.

%
%
\section{Correlation Functions}
\label{sec3}

%
%
\subsection{Leading--Order Contribution}
\label{sec31}

We turn to the evaluation of higher-order 
corrections to the imaginary part of 
correlation functions for negative $g$, and thus,
to the calculation of subleading corrections 
to the factorial growth of perturbative coefficients.
The perturbative contribution exists for 
positive and negative coupling $g$;
the cut across the negative $g$ axis is dominated by the 
instanton solution. The generating functional ${\calZ}(J)$ 
of the correlation functions is given by
\begin{equation}
\label{defZcorr}
{\calZ}(\undJ) = \frac{1}{\calN} 
\int \left[ \dd \undq(t) \right] \,
\exp\left[- \calS[\undq(t)] + 
\int\dd t\, \undJ(t) \cdot \undq(t)\right] \,,
\end{equation}
where both $\undJ(t)$ as well as $\undq(t)$ are $N$-vectors.
Note that ${\calZ}(\undJ)$ is not to be confused with 
the partition function ${\calZ}(\beta)$.
It is normalized so that,
in leading order in $g$, 
and expanded about the Gaussian saddle point,
one has ${\calZ}(\underline 0) \to 1$, i.e.,
\begin{subequations}
\begin{align}
\calN =& \; \int \left[ \dd \undq(t) \right] \,
\exp\left[-\calS_0[\undq(t)] \right] \,,
\\[0.1133ex]
{\calS}_0[\und q(t)] =& \; \int \dd t \, \left[ 
\frac12 \, \left( \frac{\partial \und q(t)}{\partial t} \right)^2 +
\frac12 \, \und q^2(t) \right] \,.
\end{align}
\end{subequations}
At leading order in the instanton contribution, the generating functional 
${\calZ}(J)$ is the sum of a 
perturbative expansion $\calZ_0$ (about the Gaussian saddle point) and an 
imaginary, exponentially small contribution
$\calZ_1$ for $g\to0$, which consists of the instanton contribution 
proportional to $\ee^{A/g}$, for $g \to 0^-$,
\begin{subequations}
\begin{align}
\label{W01}
{\calZ}(\undJ)=& \; {\calZ}_0(\undJ)+ {\calZ}_1(\undJ) \,,
\\[0.1133ex]
{\mathcal W}(\undJ) = & \;
\ln {\calZ}(\undJ) = \ln {\calZ}_0(\undJ) +
\frac{{\calZ}_1(\undJ)}{{\calZ}_0(\undJ)} 
\nonumber\\[0.1133ex]
=& \; {\mathcal W}_0(\undJ) + {\mathcal W}_1(\undJ) \,,
\end{align}
\end{subequations}
where we note that ${\calZ}_1(\undJ) $ is exponentially 
suppressed for $g \to 0^-$.
We note 
the implicit definitions ${\mathcal W}_0(\undJ) = \ln {\calZ}_0(\undJ)$
and ${\mathcal W}_1(\undJ) = {\calZ}_1(\undJ) / {\calZ}_0(\undJ)$.
The perturbative expansion defines ${\calZ}_0(\undJ)$
and holds irrespective of the sign of the coupling $g$.
By contrast, the instanton contribution ${\calZ}_1(\undJ)$
is present only for negative $g$, and this is the 
implicit assumption on which all considerations reported 
in the current section are based. We investigate 
the connected $n$-point correlation functions
$\calW^{(n)}_{A, \{ \alpha_i \}_{i=1}^n }$
and the complete $n$-point correlation functions
$\calZ^{(n)}_{A, \{ \alpha_i \}_{i=1}^n }$
for $A = 0$ (perturbative contributions)
and $A = 1$ (nonperturbative terms),
\begin{subequations}
\begin{align}
\calW^{(n)}_{A, \{ \alpha_i \}_{i=1}^n }(t_1, \ldots, t_n) &=
\left.
\left(\prod_{i=1}^n 
\frac{\delta}{\delta J_{\alpha_i}(t_i)} \right) \, 
{\mathcal W}_A(\undJ) \right|_{\undJ=\undzero} \,,
\\[2ex]
\calZ^{(n)}_{A, \{ \alpha_i \}_{i=1}^n }(t_1, \ldots, t_n) &=
\left.
\left(\prod_{i=1}^n
\frac{\delta}{\delta J_{\alpha_i}(t_i)} \right) \,
{\mathcal Z}_A(\undJ) \right|_{\undJ=\undzero} \,,
\end{align}
\end{subequations}
In order
to simplify the explicit expressions, we now assume that 
${\calS}(\undq) = {\calS}(-\undq)$ and thus, 
that correlation functions with $n$ odd vanish, 
which is certainly the case for our $\undq^4$ model.
Then, one finds, for example, for the zero-point function,
\begin{equation}
\label{einstantonsa}
{\mathcal W}_1(\undJ= \undzero) =
{\frac{ {\calZ}_1(\undzero) }{ {\calZ}_0(\undzero) }} \,, 
\qquad
{\rm Im} \, {\mathcal W}_1(\undJ=\undzero) = 
{\frac{ {\rm Im} \, {\calZ}_1(\undzero) }{ {\calZ}_0(\undzero) }} \,.
\end{equation}
For the two-point function, one finds
\begin{align}
\label{einstantonsb}
\calW^{(2)}_{1, \alpha_1 \alpha_2}(t_1, t_2) =& \;
\left.
\frac{\delta^2 }{ \delta J_{\alpha_1}(t_1) \, \delta J_{\alpha_2}(t_2)}
{\mathcal W}_1(\undJ) 
\right|_{\undJ = \undzero}
\nonumber\\[2ex]
=& \; \frac{ \calZ_{1, \alpha_1 \alpha_2}^{(2)}(t_1, t_2) }{ {\calZ}_0(\undzero)} -
\frac{ \calZ_{0, \alpha_1 \alpha_2}^{(2)}(t_1, t_2) 
{\calZ}_1(\undzero) }{ {\calZ}_0^2(\undzero) } \,, 
\end{align}
and the imaginary part of the 
two-point function is obtained as follows,
\begin{align}
\label{ImW}
\Im \calW^{(2)}_{1, \alpha_1 \alpha_2}(t_1, t_2) =& \;
\frac{ \Im \calZ_{1, \alpha_1 \alpha_2}^{(2)}(t_1, t_2) }{ {\calZ}_0(\undzero)} 
\nonumber\\[0.1133ex]
& \; - \frac{ \calZ_{0, \alpha_1 \alpha_2}^{(2)}(t_1, t_2) \, 
\Im {\calZ}_1(\undzero) }{ {\calZ}_0^2(\undzero) } \,.
\end{align}
Furthermore, we can express the imaginary 
part of the four-point function as a sum of four
terms $\calK_i$ ($i=1,\dots,4$),
\begin{align}
\label{ImW4}
& {\Im} \calW^{(4)}_{1, \alpha_1 \alpha_2 \alpha_3 \alpha_4}(t_1,t_2,t_3,t_4) =
\sum_{i=1}^4 \calK_i \,.
\end{align}
The first term involves the imaginary part 
of the four-point instanton contribution
$\Im \calZ_{1, \alpha_1 \alpha_2 \alpha_3 \alpha_4}^{(4)}$,
\begin{equation}
\calK_1 = 
\frac{ \Im \calZ_{1, \alpha_1 \alpha_2 \alpha_3 \alpha_4}^{(4)}(t_1,t_2,t_3,t_4) }%
{ {\calZ}_0(\undzero) } \,.
\end{equation}
The second term is a mixed term, involving two-point 
perturbative and two-point instanton correlation functions,
\begin{multline}
\calK_2 = 
- \frac{\calZ_{0, \alpha_1 \alpha_2}^{(2)}(t_1,t_2) \,
\Im \calZ_{1, \alpha_3 \alpha_4}^{(2)}(t_3,t_4) }{ {\calZ}_0^2(\undzero) }
\\
- \frac{\calZ_{0,\alpha_1\alpha_3}^{(2)}(t_1,t_3) \,
\Im \calZ_{1,\alpha_2\alpha_4}^{(2)}(t_2,t_4) }{ {\calZ}_0^2(\undzero) }
\\
- \frac{\calZ_{0,\alpha_1\alpha_4}^{(2)}(t_1,t_4) \,
\Im \calZ_{1, \alpha_2\alpha_3}^{(2)}(t_2,t_3) }{ {\calZ}_0^2(\undzero) } 
\\
- \frac{ \calZ_{0,\alpha_3\alpha_4}^{(2)}(t_3,t_4) \,
\Im \calZ_{1,\alpha_1\alpha_2}^{(2)}(t_1,t_2) }{ {\calZ}_0^2(\undzero) }
\\
- \frac{ \calZ_{0,\alpha_2 \alpha_4}^{(2)}(t_2,t_4) \,
\Im \calZ_{1,\alpha_1\alpha_3}^{(2)}(t_1,t_4) }{ {\calZ}_0^2(\undzero) }
\\
- \frac{ \calZ_{0,\alpha_2 \alpha_3}^{(2)}(t_2,t_3) \, 
\Im \calZ_{1,\alpha_1\alpha_4}^{(2)}(t_1,t_4) }{ {\calZ}_0^2(\undzero) } \,.
\end{multline}
The third term combines the four-point perturbative 
correlation function with the imaginary part of the 
zero-point function.
\begin{equation}
\calK_3 = 
- \frac{\calZ_{0,\alpha_1 \alpha_2 \alpha_3 \alpha_4}^{(4)}(t_1,t_2,t_3,t_4) \, 
\Im {\calZ}_1(\undzero) }{ {\calZ}_0^2(\undzero) }  \,.
\end{equation}
Note that $\Im {\calZ}_1(\undzero) $ is equal to the imaginary 
part of the partition function, up to a factor $\beta$.
Finally, the fourth term involves two perturbative
two-point functions,
\begin{multline}
\calK_4 = 
\frac{ 2 \calZ_{0,\alpha_1 \alpha_2}^{(2)}(t_1,t_2) 
\calZ_{0,\alpha_3 \alpha_4}^{(2)}(t_3,t_4) }{ {\calZ}_0^3(\undzero) }
\Im {\calZ}_1(\undzero) 
\\
+ \frac{ 2 \calZ_{0, \alpha_1 \alpha_3}^{(2)}(t_1,t_3)  
\calZ_{0, \alpha_2 \alpha_4}^{(2)}(t_2,t_4) }{ {\calZ}_0^3(\undzero) }
\Im {\calZ}_1(\undzero) 
\\
+ \frac{2 \calZ_{0,\alpha_1 \alpha_4}^{(2)}(t_1,t_4) 
\calZ_{0,\alpha_2 \alpha_3}^{(2)}(t_2,t_3) }{ {\calZ}_0^3(\undzero) } 
\Im {\calZ}_1(\undzero) \,.
\end{multline}
At leading order, $\Im \calZ_1^{(n)}$ is proportional 
to $(q_{\rm cl})^n$. The 
classical path $q_{\rm cl}$ is of order $1/\sqrt{-g}$.
So, the imaginary part 
$\Im \calZ_1^{(n)}$ of the $n$-point function
is of order $(-g)^{-n/2} \Im{\calZ}_1(\undzero)$.
This implies the inequality 
$\Im \calZ_1^{(4)} \gg \Im \calZ_1^{(2)} \gg \Im {\calZ}_1$ 
and thus, at leading order, the disconnected parts are suppressed.
Finally, a generic expression for the connected
$n$-point Green function is given by
\begin{align}
\label{sfGdef}
\sfG^{(n)}_{\{ \alpha_i \}_{i=1}^n }(t_1, \ldots, t_n) &=
\left.
\left(\prod_{i=1}^n 
\frac{\delta}{\delta J_{\alpha_i}(t_i)} \right) \, 
{\mathcal W}(\undJ) \right|_{\undJ=\undzero} \
\nonumber\\[0.1133ex]
=& \; \calW^{(n)}_{0,\{ \alpha_i \}_{i=1}^n }(t_1, \ldots, t_n) 
\nonumber\\[0.1133ex]
& \; + \calW^{(n)}_{1,\{ \alpha_i \}_{i=1}^n }(t_1, \ldots, t_n)  \,,
\end{align}
where ${\mathcal W}$ is the sum of ${\mathcal W}_0$ and 
${\mathcal W}_1$. The definition encompasses 
both the real and the imaginary part of the $n$-point
correlation function [see Eq.~\eqref{W01}].
For the imaginary part, we define
\begin{align}
\label{genimdef}
\calG^{(n)}_{\{ \alpha_i \}_{i=1}^n }(t_1, \ldots, t_n) =& \;
\Im \sfG^{(n)}_{\{ \alpha_i \}_{i=1}^n }(t_1, \ldots, t_n) 
\nonumber\\[0.1133ex]
=& \; 
\Im \calW^{(n)}_{1,\{ \alpha_i \}_{i=1}^n }(t_1, \ldots, t_n)  \,.
\end{align}

%
%
\subsection{Two--Point Correlation Function}
\label{sec32}

We investigate the (imaginary part of the) two-point Green function
$\calG_{\alpha \beta}(t_1, t_2)$, according to Eq.~\eqref{genimdef},
as follows,
\begin{align}
\label{defGab}
\calG_{\alpha \beta}(t_1, t_2) =& \;
\frac{ \Im \calZ_{1, \alpha \beta}^{(2)}(t_1, t_2) }{ {\calZ}_0(\undzero)} -
\frac{ \calZ_{0, \alpha \beta}^{(2)}(t_1, t_2) \, 
\Im {\calZ}_1(\undzero) }{ {\calZ}_0^2(\undzero) } 
\nonumber\\[0.1133ex]
=& \; [ G_{\alpha\beta}(t_1, t_2) ]_1 + 
[ G_{\alpha\beta}(t_1, t_2) ]_2 
\nonumber\\[0.1133ex]
\approx & \; [ G_{\alpha\beta}(t_1, t_2) ]_1 \,.
\end{align}
Note that in the last step of the previous expression,
we have reported only the leading term in $ g $; 
however, $ [ G_{\alpha\beta}(t_1, t_2) ]_2 $ will be important 
when we will consider the first subleading order.

It is useful to remark that all
$\calZ$ quantities are now understood in the sense of Eq.~\eqref{defZcorr},
i.e., without $\int \dd q_0$.
We need to evaluate an integral 
of the form $\oint [\dd \und q(t)] \, F[\und q(t)]$
with the help of Eq.~\eqref{FFF},
where in leading order in $g$, 
one has
\begin{align}
F[\und q(t)] =& \; q_{\cl,\alpha}(t_1) \, q_{\cl,\beta}(t_2) 
\\[0.1133ex]
\approx & \; 
-\frac{1}{g} \, u_\alpha \, u_\beta \, 
\xi_\cl(t_1 - t_0) \, \xi_\cl(t_2 - t_0) \,,
\nonumber\\[0.1133ex]
\left< F[\undq(t)] \right>_{S_{N-1}}
\approx & \; -\frac{1}{g} \, \left< u_\alpha \, u_\beta \right>_{S_{N-1}} 
\xi_\cl(t_1 - t_0) \, \xi_\cl(t_2 - t_0) 
\nonumber\\[0.1133ex]
=& \; -\frac{1}{g} \, \frac{\delta_{\alpha\beta}}{N} \, 
\xi_\cl(t_1 - t_0) \, \xi_\cl(t_2 - t_0) \,.
\end{align}
We recall that, according to the remarks surrounding 
Eq.~\eqref{FF2}, we need to supplement the collective
coordinate $t_0$ in the actual path. 
Hence, we can approximate, in leading order,
\begin{multline}
\label{ImGalphabetaLeading}
\calG_{\alpha\beta}(t_1, t_2) \approx 
\left( \frac{1}{\sqrt{2 \pi}} \right)^N \;
\frac{2 \; \pi^{N/2}}{\Gamma(N/2)} \;
\left( \frac{1}{2} \right) \, 
\\[0.1133ex]
\times \left( - \frac{ 4 }{ g } \right)^{(N-1)/2} \,
\left( - \frac{ 4 }{ 3 g } \right)^{1/2} \,
\exp \left(  \frac{4}{3 g} \right)  \,
\\[0.1133ex]
\times 
\left[ -\det \left( \frac{ 1 }{\bfM_0 } \bfM_{\LL} \right) \right]^{-1/2} \;
\left[ \det \left( \frac{ 1 }{\bfM_0 } \bfM_{\TT} \right) \right]^{-N/2} \;
\\[0.1133ex]
\times 
\left( -\frac{1}{g} \, \frac{\delta_{\alpha\beta}}{N} \, 
\int \dd t_0  \,
\xi_\cl(t_1 - t_2 - t_0) \, \xi_\cl(- t_0) \right)  \,
\\[0.1133ex]
= -\frac{1}{g} \frac{1}{ \Gamma(N/2)} 
\frac{\delta_{\alpha\beta}}{N} 
\left( - \frac{ 8 }{ g } \right)^{N/2} 
\exp \left( \frac{4}{3 g} \right) 
\frac{4 (t_1 - t_2)}{\sinh(t_1 - t_2)} .
\end{multline}
This imaginary part $\calG_{\alpha\beta}(t_1,t_2)$
is positive for negative $g$,
in contrast to the negative imaginary part of the 
ground-state resonance energy.
[We recall that, according to Eq.~\eqref{defGab},
the imaginary part of the Green function itself,
not the entire Green function, is denoted as 
$\calG_{\alpha\beta}(t_1,t_2)$.]

%
%
\subsection{Some Observations}
\label{sec33}

Before we go {\em in medias res}, four observations should be made.

{\em (i)} We are interested in the corrections
to the result~\eqref{ImGalphabetaLeading}
of relative order $g$. One of these corrections
can be obtained almost automatically,
by observing that the derivation of the leading
term given in Eq.~\eqref{ImGalphabetaLeading}
does not entail path integrals 
except in leading order; the 
product of the two classical field configurations
$\int \dd t_0  \, \xi_\cl(t_1 - t_2 - t_0) \, \xi_\cl(- t_0) $
simply drops out as a prefactor of the integral.
Hence, the two-point correlation function
receives the same relative correction as the 
partition function itself; i.e., it has to be 
multiplied by [see Eq.~\eqref{Z1corr}]
\begin{align}
\label{defFZ}
\calF_\calZ =& \;
\left( \frac{1}{\beta} \frac{ {\rm Im} \calZ(\beta) }{ \calZ_0(\beta) } \right) \,
\left[ \left.
\left( \frac{1}{\beta} \frac{ {\rm Im} \calZ(\beta) }{ \calZ_0(\beta) } \right) 
\right|_0 \right]^{-1} 
\nonumber\\[0.1133ex]
=& \;
1 + g \, \left( \frac{7}{32} N^2 + \frac{9}{16} N + \frac{5}{24} \right) \,,
\end{align}
where, with the subscript zero, 
we denote the leading contribution to the imaginary part
of the ground-state energy.
We note that in the two-point correlation function,
the integration over the collective coordinate $\int \dd t_0$ 
is carried out over the arguments of the 
classical field configuration;
in the derivation of the partition function,
by contrast,
it leads to a factor $\int \dd t = \beta$,
which is later divided out in calculating the 
energy. We note that the leading term in
the two-point correlation function
can be written as follows,
\begin{multline}
\calG_{\alpha\beta}(t_1, t_2) \approx 
\left. \left( \frac{1}{\beta} \;
\frac{{\rm Im} \; \calZ_1(\beta)}{\calZ_{0}(\beta)} \right) \right|_0 
\\[0.1133ex]
\times 
\left( -\frac{1}{g} \, \frac{\delta_{\alpha\beta}}{N} \, \int \dd t_0  \,
\xi_\cl(t_1 - t_2 - t_0) \, \xi_\cl(- t_0) \right) \,.
\end{multline}
Replacing the prefactor according to 
\begin{equation}
\label{repl}
\left. \left( \frac{1}{\beta} \frac{ {\rm Im} \calZ(\beta) }{ \calZ_0(\beta) } \right) 
\right|_0 \to
\frac{1}{\beta} \frac{ {\rm Im} \calZ(\beta) }{ \calZ_0(\beta) } 
\end{equation}
takes care of the correction,
and that replacement exactly amount to the 
multiplication of the leading-order result by the 
correction factor $\calF_\calZ$.

{\em (ii)} The angular symmetry of the problem implies that 
\begin{equation}
\label{defSCALARG1}
\calG_{\alpha\beta}(t_1, t_2) = 
\frac{\delta_{\alpha\beta}}{N} \, 
\calG_{\gamma\gamma}(t_1, t_2) \,.
\end{equation}
Hence, we can restrict the discussion, in the 
following, to the function
\begin{equation}
\label{defSCALARG2}
\calG(t_1, t_2) = \calG_{\gamma\gamma}(t_1, t_2) 
= \delta_{\alpha\beta} \, \calG_{\alpha\beta}(t_1, t_2) \,,
\end{equation}
an operation which also eliminates the necessity to 
do angular averaging.

{\em (iii)} We recall the action factor 
$F_1$ from Eq.~\eqref{factorF1},
the Jacobian factor $F_2$ from Eq.~\eqref{factorF2},
and the perturbative factor $F_3$ from Eq.~\eqref{factorF3}.
The correction due to the factor $F_3$,
in relative order $g$,
is already taken into account 
in the denominator of the 
perturbative partition function $\calZ_0(\beta)$ in 
the replacement in Eq.~\eqref{repl}.
If we are thinking about the calculation of 
perturbative corrections about the instanton saddle 
point of the two-point 
correlation function, then we must consider that the 
leading term is proportional to 
\begin{equation}
\label{leadingqq}
\left< q_{\cl, \alpha}(t_1 - t_0) \, q_{\cl,\beta}(t_2 - t_0) 
\right>_{S_{N-1}}  \,,
\end{equation}
according to Eq.~\eqref{ImGalphabetaLeading}.
Corrections of relative order $g$ are obtained 
in two ways, first, by replacing, in Eq.~\eqref{leadingqq},
both classical paths by fluctuations,
which results in a term of relative order $g$,
because the fluctuations are of order $g^0$,
while the classical field configurations are 
of order $1/\sqrt{-g}$.
In this case, the calculation proceeds simply by 
evaluating the path integral, without any further 
perturbative corrections from either $F_1$ or $F_2$,
and is already of the required relative order $g$.

The second way to obtain a correction of 
relative order $g$ is to replace only one of the
classical field configurations in Eq.~\eqref{leadingqq}
by a fluctuation, and to contract the remaining term
with the term $\calF_\calJ$ which contains the terms
up to relative order $\sqrt{-g}$ from the 
product $F_1 \, F_2$, and reads as follows,
\begin{align}
\label{defFJ}
\calF_\calJ =&\; \left. F_1 \, F_2 \right|_{\sqrt{-g}} 
= 1 + \sqrt{-g} \, \left[
\int \dd t \und \xi_{\rm cl}(t) \cdot \und\chi(t) \;\; \und \chi^2(t)
\right.
\nonumber\\[0.1133ex]
& \; \left.
+ \frac{3}{4} \int \dd t \, \dot{\und \chi}(t) \cdot 
\dot{\und \xi}_{\rm cl}(t)
+ \frac{N-1}{4} 
\int\dd t\,{\underline \chi}(t)\cdot {\underline \xi}_{\rm cl}(t)
\right] \,.
\end{align}
Finally, the third way to obtain a correction to the 
two-point function is via the perturbative subtraction term 
[the second term in Eq.~\eqref{ImW}],
which involves the perturbative (Gaussian) correlation
function $\calZ_{0, \alpha_1 \alpha_2}^{(2)}(t_1, t_2)$.
It is somewhat analogous to the factor $F_3$ for the partition function.

{\em (iv)} As it will turn out, 
one can actually show that the two-point Green function 
$G(t_1, t_2)$ is a function of the time difference
$t_1 - t_2$.
This has consequences for the evaluation of the
two-point correlator at zero momentum transfer,
as follows. Namely, {\em a priori}, one would formulate the 
Fourier transform of the two-point correlator as follows,
\begin{equation}
\calG(p_1, p_2) = 
\int \dd t_1 \, \int \dd t_2 \, 
\ee^{ - \ii p_1 t_1 - \ii p_2 t_2 } \,
\calG(t_1, t_2) \,.
\end{equation}
Using the property
\begin{equation}
\label{GDIFF}
\calG(t_1, t_2) = \calG(t_1 - t_2) \,,
\end{equation}
one finds
\begin{equation}
\calG(p_1, p_2) = 
2\pi \, \delta(p_1 + p_2) \, \calG(p_1) \,.
\end{equation}
We will be interested here in the 
two-point correlation function
at zero momentum transfer, which,
in view of the above considerations,
is just the Fourier transform of $G(\tau)$
at zero momentum, i.e.,
\begin{align}
\label{G0}
\calG(p = 0) =& \; \int \dd \tau \, \calG(\tau) \,,
\\[0.1133ex]
\label{GP0}
\frac{\partial^2}{\partial p^2} \calG(p = 0) =& \; 
\int \dd \tau \, (-\tau^2) \, \calG(\tau) \,,
\end{align}
where the latter expression enters the Callan--Symanzik
equation.

%
%
\subsection{First Correction Term}
\label{sec34}

Let us summarize the formulas mentioned above.
We have, for the first term in Eq.~\eqref{defGab},
\begin{equation}
[ \calG_{\alpha\beta}(t_1, t_2) ]_1 =
\frac{\delta_{\alpha\beta}}{N} \,
[ \calG_{\gamma\gamma}(t_1, t_2) ]_1 =
\frac{\delta_{\alpha\beta}}{N} \,
[ \calG(t_1, t_2) ]_1 \,.
\end{equation}
For later reference, it is customary to define
a recurrent prefactor as
\begin{equation}
\label{defcalQ}
\calQ(g) = \frac{1}{\Gamma(N/2)} \, 
\left( - \frac{8}{g} \right)^{N/2} \,
\exp\left( \frac{4}{3 \, g} \right) \,.
\end{equation}
Corrections to the two-point function can be 
derived based on Eqs.~\eqref{defFZ} and~\eqref{defFJ},
and lead to the formula
\begin{multline}
\label{masterZ}
\frac{\left[ \calG(t_1, t_2) \right]_1 }{\calQ(g)}
= \int \dd t_0 
\left< \calF_\calJ \; q_\gamma(t_1 - t_0) \, q_\gamma(t_2 - t_0) 
\right> \, \calF_\calZ
\\[0.1133ex]
\approx \int \dd t_0 
\left< q_{\cl,\gamma}(t_1 - t_0) \, q_{\cl,\gamma}(t_2 - t_0) 
\right> \, \calF_\calZ
\\[0.1133ex]
+ \int \dd t_0 
\left< (\calF_\calJ - 1) \; q_\gamma(t_1 - t_0) \, q_\gamma(t_2 - t_0) 
\right>,
\end{multline}
which is valid up to relative order $g$.
Here, the path integral expectation value $\left< \cdot \right>$
has been defined in Eq.~\eqref{WEIGHTLT}.
The first term has the classical field configuration
$q_{\cl,\gamma}(t_1 - t_0)$
and the correction factor $\calF_\calZ$,
while the second has the 
correction factor $\calF_\calJ$.
We can thus write the two-point correlation function,
up to relative order $g$, as follows,
\begin{multline}
\label{interim}
\frac{\left[ \calG(t_1, t_2) \right]_1 }{\calQ(g)} =
\int \dd t_0 \, \left( - \frac{1}{g} \right) 
\xi_{\rm cl}(t_1 - t_0) \, \xi_{\rm cl}(t_2 - t_0)  \,
\calF_\calZ
\\[0.1133ex]
+ \int \dd t_0 \,
\left< \chi_{\LL}(t_1 - t_0) \, \chi_{\LL}(t_2 - t_0) 
\right.
\\[0.1133ex]
\left. 
+ \und \chi_{\TT}(t_1 - t_0) \cdot \und \chi_{\TT}(t_2 - t_0)
\right> 
\\[0.1133ex]
+ \int \dd t_0 \, \left< 
\sqrt{- \frac{1}{g} } \, \chi_{\LL}(t_1 - t_0) \, 
\xi_{\rm cl}(t_2 - t_0) \; 
\left( \calF_\calJ - 1 \right)
\right>
\\[0.1133ex]
+ \int \dd t_0 \, \left< 
\sqrt{- \frac{1}{g} } \, \chi_{\LL}(t_2 - t_0) \,
\xi_{\rm cl}(t_1 - t_0) \, 
\left( \calF_\calJ - 1 \right) \right> \,,
\end{multline}
where the first and the second term have already been treated.
The Wick theorem immediately leads to 
\begin{multline}
\label{trivialid}
\left< \chi_{\LL}(t_1 - t_0) \, \chi_{\LL}(t_2 - t_0)
+ \und \chi_{\TT}(t_1 - t_0) \cdot \und \chi_{\TT}(t_2 - t_0)
\right>
\\
= \bfDelta_{\LL}(t_1 - t_0, t_2 - t_0)
+ (N - 1) \, \bfDelta_{\TT}(t_1 - t_0, t_2 - t_0) \,.
\end{multline}
For the third and the fourth term in Eq.~\eqref{interim},
one consults the definition of
$\calF_\calJ$ and applies the Wick theorem in order,
to obtain
\begin{multline}
\label{interim2}
\frac{[ \calG(t_1, t_2) ]_1}{\calQ(g)} =
\int \dd t_0 \left\{ - \frac{1}{g} 
\xi_{\rm cl}(t_1 - t_0) \, \xi_{\rm cl}(t_2 - t_0)  
\calF_\calZ
\right.
\\[0.1133ex]
\left. + \left[ \bfDelta_{\LL}(t_1 - t_0, t_2 - t_0)
+ (N - 1) \bfDelta_{\TT}(t_1 - t_0, t_2 - t_0) \right] \right\}
\\[0.1133ex]
+ \biggl\{ \int \dd t_0 \, \int \dd t \,
\xi_{\rm cl}(t_1 - t_0) \, \xi_{\rm cl}(t) \,
\left[ 3 \, \bfDelta_{\LL}(t_2 -t_0, t) \, \bfDelta_{\LL}(t, t)
\right.
\\[0.1133ex]
\left.
+ (N-1) \, \bfDelta_{\LL}(t_2 -t_0, t) \, \bfDelta_{\TT}(t, t) \right]
\\[0.1133ex]
- \frac34 \int \dd t_0 \int \dd t \, 
\xi_{\rm cl}(t_1 - t_0) \, {\ddot \xi}_{\rm cl}(t) \,
\bfDelta_{\LL}(t_2 - t_0, t) 
\\[0.1133ex]
+ \frac{N-1}{4} \int \dd t_0 \int\dd t\, 
\xi_{\rm cl}(t_1 - t_0) \, \xi_{\rm cl}(t) \, 
\bfDelta_{\LL}(t_2 - t_0, t)  \biggr\}
\\[0.1133ex]
+ \{ t_1 \leftrightarrow t_2 \} \,,
\end{multline}
where $\{ t_1 \leftrightarrow t_2 \}$ denotes the 
terms listed in the previous curly brackets,
with the time variable $t_1$ and $t_2$ interchanged.
The first two terms do not involve an additional integration over 
$t$, because they do not incur corrections from the 
factor $\calF_\calJ$.
In the result, we have the integration over $t$, 
from the product $F_1 \, F_2$,
the integration over $t_0$, from the collective coordinate,
and the integration over $t'$, from the evaluation of the
correlation function at zero momentum transfer.

If we shift, in Eq.~\eqref{interim2},
the integration variable, uniformly, according to
\begin{equation}
\label{shift1}
t_0 \to t_0 + t_2 \,,
\end{equation}
then we can show the time translation invariance identity
[see Eq.~\eqref{GDIFF}]
\begin{equation}
[ \calG(t_1, t_2) ]_1 =
[ \calG(t_1 - t_2, 0) ]_1 =
[ \calG(t')]_1 \,,
\quad
t' = t_1 - t_2 \,.
\end{equation}
However, in the last term in Eq.~\eqref{interim2}
[the one characterized by the replacement
$(t_1 \leftrightarrow t_2)$],
it is actually advantageous to shift the integration variable 
according to $t_0 \to t_0 + t_1$,
at variance with Eq.~\eqref{shift1}.

Eventually, we will need to calculate the integral
[see Eq.~\eqref{G0}]
\begin{equation}
\int \dd t' \, [ \calG(t') ]_1 = [ \calG(p=0) ]_1 \,.
\label{G2ZERODEF}
\end{equation}
We find
\begin{multline}
\frac{[ \calG(p = 0) ]_1}{\calQ(g)} =
\int \dd t' \, \int \dd t_0 \, 
\biggl\{ - \frac{1}{g} 
\xi_{\rm cl}(t' - t_0) \, \xi_{\rm cl}(- t_0) 
\calF_\calZ
\\[0.1133ex]
+ \left[ \bfDelta_{\LL}(t' - t_0, - t_0)
+ (N - 1) \, \bfDelta_{\TT}(t' - t_0, - t_0) \right]
\biggr\}
\\[0.1133ex]
+ 2 \int \dd t' \, \int \dd t_0 \, \int \dd t \,
\biggl\{ \xi_{\rm cl}(t' - t_0) \, \xi_{\rm cl}(t) 
\\[0.1133ex]
\times \left[ 3 \, \bfDelta_{\LL}(-t_0, t) \, \bfDelta_{\LL}(t, t)
+ (N-1) \, \bfDelta_{\LL}(-t_0, t) \, \bfDelta_{\TT}(t, t) \right]
\\[0.1133ex]
- \frac34 \xi_{\rm cl}(t' - t_0) \, {\ddot \xi}_{\rm cl}(t) \,
\bfDelta_{\LL}(- t_0, t) 
\\[0.1133ex]
+ \frac{N-1}{4} \xi_{\rm cl}(t' - t_0) \, \xi_{\rm cl}(t) \, 
\bfDelta_{\LL}(- t_0, t) \biggr\} \,.
\label{interim4}
\end{multline}
This integral is divergent for large $\beta$, 
but the infinities are removed upon consideration of the 
perturbative term $[ \calG(p=0) ]_2$, to be considered in the following.

In the last term in Eq.~\eqref{interim4}, we have
the integration over $t$, which is the
integration variable in the perturbative factor $\calF_\calJ$,
the integration over the collective coordinate $t_0$,
and the integration over the time translation variable $t' = t_1 - t_2$
of the Green function.
The presence of the instanton, which vanishes exponentially
for large argument, guarantees the convergence of the integral.

%
%
\subsection{Second Correction Term}
\label{sec35}

We now concentrate on the second term in Eq.~\eqref{defGab},
\begin{align}
\label{G2}
[ \calG_{\alpha\beta}(t_1, t_2) ]_2 =
- \frac{ \calZ_{0, \alpha \beta}^{(2)}(t_1, t_2) \, 
\Im {\calZ}_1(\undzero) }{ {\calZ}_0^2(\undzero) } 
\end{align}
and we can remember that, according to our previous 
considerations, the discussion can be restricted to the 
expression
\begin{align}
[ \calG(t_1, t_2) ]_2 =
[ \calG_{\gamma\gamma}(t_1, t_2) ]_2 =
- \frac{ \calZ_{0, \gamma\gamma}^{(2)}(t_1, t_2) }{ {\calZ}_0(\undzero) }
\; \frac{ \Im {\calZ}_1(\undzero) }{ {\calZ}_0(\undzero) } \,.
\end{align}
In comparison to $[ \calG(t_1, t_2) ]_1$, the expression
$[ \calG(t_1, t_2) ]_2$ is of relative order $g$,
because it lacks the presence of the 
classical paths, which are, themselves, of order $\sqrt{-1/g}$.
So, we evaluate the expression~\eqref{G2} to 
leading order only.

We have already anticipated the cancelation mechanism for the 
$\beta$ parameter; 
indeed, there is no dependence on $\beta$ in
$\calZ_{0,\alpha\beta}^{(2)}(t_1, t_2)$; however,
there is a multiplicative factor $\beta$ in ${\rm Im} \, \calZ_1 $,
due to the integral over the collective coordinate $t_0$.
This factor $\beta$ should cancel against 
other divergences in $\beta$, to be found 
in the integrals over the propagators in 
$\calZ_{1,\alpha\beta}^{(2)}(t_1, t_2)$.
In fact, we show in the following, 
that the term $[ \calG(t_1, t_2) ]_2$ exactly furnishes
the terms necessary for the removal of the infinities
in Eq.~\eqref{interim4}.

First, we have, upon perturbative expansion,
\begin{align}
\label{ZGalphabeta}
\frac{\calZ_{0,\gamma\gamma}^{(2)}(t_1, t_2)}%
{\calZ_0(\undzero)} = & \;
\left< q_\gamma(t_1) \, q_\gamma(t_2) \right>_0
= N \, \bfDelta_0(t_1, t_2) \,,
\end{align}
where the integration measure $\langle \cdot \rangle_0$ 
has been defined in Eq.~\eqref{WEIGHT0_ON}.
Compared to the instanton 
partition function $\calZ_1(\beta)$, the 
generating function $\calZ_1(\undJ)$ lacks the 
integration over the end point $q_0$, 
while the same is true for the 
perturbative contributions
$\calZ_0(\beta)$ versus $\calZ_0(\undJ)$.
However, the lacking integration cancels
in the ratio, and we can write,
with Eq.~\eqref{LeadQuartONInst},
\begin{align}
\frac{{\rm Im} \; \calZ_1(\undzero)}{\calZ_{0}(\undzero)} =& \;
\frac{{\rm Im} \; \calZ_1(\beta)}{\calZ_{0}(\beta)} =
\frac{\beta}{ \Gamma(N/2)} \; \left( - \frac{ 8 }{ g } \right)^{N/2} \,
\exp \left( \frac{4}{3 g} \right) 
\nonumber\\[0.1133ex]
& \; \times
\left[ 1 + g \, \left( \frac{7}{32} N^2 + \frac{9}{16} N + \frac{5}{24} \right)
\right] \,.
\end{align}
So, we finally get 
\begin{align}
[ \calG(t_1, t_2) ]_2 =& \;
-\frac{1}{ \Gamma(N/2)} \; \left( - \frac{ 8 }{ g } \right)^{N/2} \,
\exp \left( \frac{4}{3 g} \right) 
\nonumber\\[0.1133ex]
& \; \times \left[ N \, \beta \, \bfDelta_0(t_1, t_2) \right] \,.
\end{align}
Here, according to Eq.~\eqref{Delta0}, the free propagator
$\bfDelta_0(t_1, t_2)$ has the translation invariance property
\begin{equation}
\bfDelta_0(t_1, t_2) = 
\bfDelta_0(t_1 - t_2) =
\bfDelta_0(t_1 - t_2, 0) \,.
\end{equation}
For the purposes of the removal of the infinities 
discussed in Sec.~\ref{sec35},
we can reformulate the Fourier transform of 
this expression at zero momentum as follows,
\begin{multline}
\label{interim5}
[ \calG(p=0) ]_2 =
-\frac{1}{ \Gamma(N/2)} \; \left( - \frac{ 8 }{ g } \right)^{N/2} \,
\exp \left( \frac{4}{3 g} \right) \,
\\[0.1133ex]
\times \int \dd t_0 \int \dd t'
\left[ \bfDelta_0(0, t') + (N-1) \, \bfDelta_0(0, t') \right] \,,
\end{multline}
where $t' = t_1 - t_2$.
We have replaced $\beta \to \int \dd t_0$.
We recall that the integration limits in 
all given integrals cover the range 
$-\infty < \tau < \infty$ for all 
Euclidean time parameters $\tau$,
unless indicated otherwise;
however, these limits are incurred
in terms of the limiting process
$-\beta/2 < \tau < \beta/2$,
where we let $\beta \to \infty$.

\begin{figure}[t!]
\begin{center}
\begin{minipage}{1.0\linewidth}
\begin{center}
\includegraphics[width=0.95\linewidth]{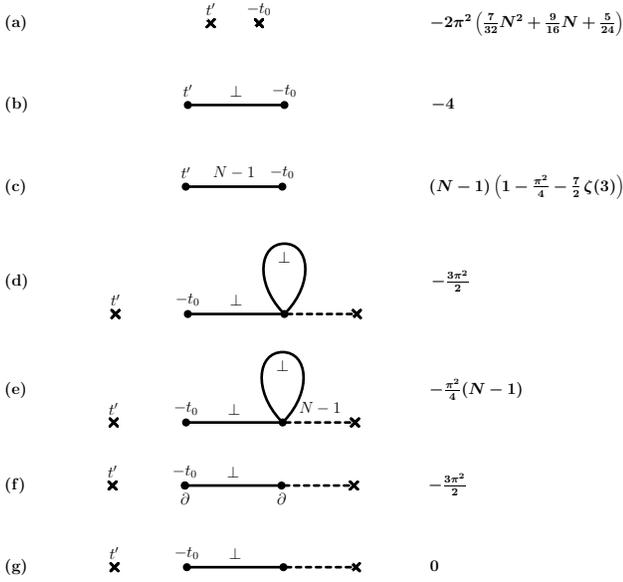}
\hfill
\caption{\label{figg3} Diagrammatic representation
of the seven two-loop corrections to the 
two-point function at zero momentum, 
for a one-dimensional $\phi^4$ theory
with an $O(N)$ internal symmetry.
The contribution of the diagrams is written beside each
contribution. The total result of order $g$ is
of the form given in Eq.~\eqref{GP0RES}.
One of the classical field configurations,
associated with the time variable $t'$
(or $t'-t_0$, before a suitable change of variable),
is somehow ``detached'' from the rest of the diagram.
Recall that the transverse character of the 
propagator (excluding the instanton configurations)
is denoted by the symbol $\perp$, and
that the variable $t' = t_1 - t_2$ 
enters in view of the time translation invariance
of the Green function.
Incidentally, the diagrams for the 
second derivative of the two-point function
(Sec.~\ref{sec37}),
and for the four-point function
(Sec.~\ref{sec38}),
are the same as those depicted here, 
with (in the case of the four-point function)
two more detached instantons.}
\end{center}
\end{minipage}
\end{center}
\end{figure}

%
%
\subsection{Evaluation of the Corrections}
\label{sec36}

We add the expressions from Eqs.~\eqref{interim4} and~\eqref{interim5}
and consider the sum of $[ \calG(p=0) ]_1$ and $[ \calG(p=0) ]_2$.
The substitutions $t' \to t' + t_0$, and subsequently $t_0 \to -t_0$
serve to simplify the expressions
(the Jacobian in each case is unity).
One can finally write $\calG(p = 0)$ as
\begin{equation}
\frac{\calG(p = 0)}{\calQ(g)} =
\sum_{i=1}^7 \calR_i \,,
\label{interim7} 
\end{equation}
where the $\calR_i$ terms ($i=1,\dots,7$) are defined 
in the following (see also Fig.~\ref{figg3}).
For the evaluation of the expression of $\calR_1$,
we refer to integral $H_1$ listed in Appendix~\ref{appendixb},
and write
\begin{align}
\calR_1 =& \; 
- \frac{1}{g} \, \left[ \int \dd t' \, \xi_{\rm cl}(t') \right]^2  \calF_\calZ
= -\frac{1}{g} (H_1)^2 \calF_Z 
\nonumber\\[0.1133ex]
=& \; -\frac{2 \pi^2}{g}
- 2 \pi^2 \left( \frac{7}{32} N^2 + \frac{9}{16} N + \frac{5}{24} \right) \,,
\end{align}
where we use Eq.~\eqref{defFZ}.
For the term $\calR_2$, one has
\begin{equation}
\calR_2 = \int \dd t' \, \int \dd t_0 \,
\left[ \bfDelta_{\LL}(t', t_0) - \bfDelta_0(t', t_0) \right] = I_1 = -4 \,.
\end{equation}
For the integral $I_1$, we again refer to Appendix~\ref{appendixb}.
The integral $\calR_3$ involves the transverse propagator, 
\begin{align}
\calR_3 =& \; (N - 1) \int \dd t' \, \int \dd t_0 \,
\left[ \bfDelta_{\TT}(t', t_0) - \bfDelta_0(t', t_0) \right] 
\nonumber\\[0.1133ex]
=& \; (N-1) I_2 =
(N-1) \left[ 1 - \frac{\pi^2}{4} - \frac72 \, \zeta(3) \right] 
\end{align}
(see also the Appendix~\ref{appendixb}).
The rest of the terms are
\begin{subequations}
\begin{align}
\calR_4 =& \; 6 \int \dd t' \, \xi_{\rm cl}(t') \, 
\int \dd t \, \xi_{\rm cl}(t) \, \bfDelta_{\LL}(t, t) \, 
\int \dd t_0 \, \bfDelta_{\LL}(t, t_0) 
\nonumber\\[0.1133ex]
=& \; 6 \, H_1 \, J_3 = \frac{3 \pi^2}{2} \,,
\\[0.1133ex]
\calR_5 =& \; 2 (N-1) \int \dd t' \, \xi_{\rm cl}(t') \, 
\int \dd t \, \xi_{\rm cl}(t) \, \bfDelta_{\TT}(t, t) \, 
\\[0.1133ex]
& \; \times \int \dd t_0 \bfDelta_{\LL}(t, t_0) = 2 (N-1) H_1 J_4 
= \frac{\pi^2}{4} (N-1) \,,
\nonumber\\[0.1133ex]
\calR_6 =& \; - \frac32 \int \dd t' \, \xi_{\rm cl}(t') \, 
\int \dd t \, {\ddot \xi}_{\rm cl}(t) \,
\int \dd t_0 \, \bfDelta_{\LL}(t, t_0) 
\nonumber\\[0.1133ex]
=& \; - \frac32 \, H_1 \, J_2 
= - \frac{3 \pi^2}{2} \,,
\\[0.1133ex]
\calR_7 =& \; \frac{N-1}{2} \int \dd t' \, \xi_{\rm cl}(t') \, 
\int \dd t \, \xi_{\rm cl}(t) \, \int \dd t_0 \, \bfDelta_{\LL}(t, t_0) 
\nonumber\\[0.1133ex]
=& \; \frac{N-1}{2} \, H_1 J_1 = 0 \,.
\end{align}
\end{subequations}
The integrals $J_i$ ($i = 1,\dots, 4$) are listed in 
Appendix~\ref{appendixb}. The end result is 
\begin{align}
\label{GP0RES}
\frac{\calG(p = 0)}{\calQ(g)} =& \;
- \frac{2 \pi^2}{g} + 
\left( \frac72 \zeta(3) - 5 - \frac{5 \pi^2}{12} \right)
\nonumber\\[0.1133ex]
& \; \left[ 1 - \frac{9 \pi^2}{8} - \frac72 \zeta(3) \right] \, N
- \frac{7 \pi^2}{16}  \, N^2 \,.
\end{align}
It is interesting to note that, in the limit 
$N \to \infty$, the leading contribution to the 
coefficient of relative order $g$ comes from the 
partition function correction $\calF_\calZ$.
Furthermore, in the limit of small $g$,
the imaginary part of the Green
function described by $\calG(p = 0)$ is positive.

%
%
\subsection{Second Derivative of the Correlator}
\label{sec37}

In order to evaluate the second derivative of 
the two-point correlation function,
we recall Eq.~\eqref{interim4},
subtract the perturbative term with subscript ``2'',
and insert a factor 
$(-t'^2)$ in the $t'$ integration.
\begin{equation}
\label{GGP_SEVEN}
\frac{ \frac{\partial^2}{\partial p^2} \calG(p = 0) }{\calQ(g)} =
\sum_{i=1}^7 \calS_i \,.
\end{equation}
After appropriate substitutions in the integration 
variables, we obtain the following expression,
the following integrals are generated,
after obvious symmetry considerations,
\begin{align}
\calS_1 =& \; 
\frac{2}{g} \int \dd t' \, t'^2 \, \xi_{\rm cl}(t') \, 
\int \dd t_0 \, \xi_{\rm cl}(t_0) \calF_\calZ =
\frac{2}{g} \, H_3 \, H_1 \, \calF_\calZ 
\nonumber\\[0.1133ex]
= & \; \frac{\pi^4}{g} + \pi^4 \,
\left( \frac{7}{32} N^2 + \frac{9}{16} N + \frac{5}{24} \right) \,.
\end{align}
The subtracted propagators, with the momentum 
derivative insertion, give rise to the following 
expressions,
\begin{align}
\calS_2 =& \; \int \dd t' \, \int \dd t_0 \, 
[-(t' - t_0)^2] \, 
\left[ \bfDelta_\LL(t', t_0) -
\bfDelta_0(t', t_0) \right] 
\nonumber\\[0.1133ex]
=& \; -{\overline K}_1 = -2 - \frac{\pi^2}{2} - 21 \, \zeta(3) \,,
\\[0.1133ex]
\calS_3 =& \; (N-1) \int \dd t' \, \int \dd t_0 \, 
[-(t' - t_0)^2] 
\nonumber\\[0.1133ex]
& \; \times \left[ \bfDelta_\TT(t', t_0) -
\bfDelta_0(t', t_0) \right] 
\nonumber\\[0.1133ex]
=& \; -(N-1) {\overline K}_2 = 
(N - 1) \left( -6 + \frac{\pi^4}{8} + \frac{93}{2} \, \zeta(5) \right) \,.
\end{align}
The ${\overline K}$ integrals are listed in Appendix~\ref{appendixb}.
The rest of the terms involve instanton configurations.
The first of these is
\begin{align}
\calS_4 =& \; 6 \int \dd t' \, \int \dd t_0 \, \int \dd t \,
[-(t' - t_0)^2] \, 
\xi_{\rm cl}(t') \, \xi_{\rm cl}(t) \,
\nonumber\\
& \; \times \bfDelta_{\LL}(t_0, t) \, \bfDelta_{\LL}(t, t)
\nonumber\\
=& \; - 6 H_3 \, J_3 - 6 H_1 \, L_3 = 
+\frac{\pi^2}{2} - \frac{9 \pi^4}{4} \,.
\end{align}
The term with combined transverse and longitudinal 
propagators is
\begin{align}
\calS_5 =& 2 (N-1) \, \int \dd t' \, 
\int \dd t_0 \, 
\int \dd t \,
[-(t' - t_0)^2] \, 
\xi_{\rm cl}(t') \, \xi_{\rm cl}(t) 
\nonumber\\
& \; \times \bfDelta_{\LL}(t_0, t) \, \bfDelta_{\TT}(t, t)
\nonumber\\
=& \; -2 (N-1) \, ( H_3 \, J_4 + H_1 \, L_4 ) 
= -\frac{5 \pi^4}{8} (N-1) \,.
\end{align}
The term with the second derivative of the instanton is
\begin{align}
\calS_6 =& \; \frac32
\int \dd t' \, \int \dd t_0 \, \int \dd t \,
(t' - t_0)^2 \, 
\xi_{\rm cl}(t') \, {\ddot \xi}_{\rm cl}(t) \,
\bfDelta_{\LL}(t_0, t) 
\nonumber\\
=& \; \frac32 ( H_3 \, J_2 + H_1 \, L_2 ) 
= \frac{3 \pi^4}{2} \,.
\end{align}
The last term generated by the Jacobian factor $\calF_\calJ$ is
\begin{align}
\calS_7 =& \; -\frac{N-1}{2} 
\int \dd t' \int \dd t_0 \int \dd t (t' - t_0)^2  
\nonumber\\
& \; \times \xi_{\rm cl}(t') \xi_{\rm cl}(t) \,
\bfDelta_{\LL}(t_0, t) 
\nonumber\\
= & \; -\frac{N-1}{2} ( H_3 \, J_1 + H_1 \, L_1 ) = 
-\frac{\pi^4}{4} \, (N-1) \,.
\end{align}
The overall result is 
\begin{align}
\label{GPP0}
\frac{\frac{\partial^2}{\partial p^2} \calG(p = 0)}{\calQ(g)} =& \;
\frac{\pi^4}{g} +
\frac{5 \pi^4}{24} + 4 
- 21 \, \zeta(3) - \frac{93}{2} \, \zeta(5) 
\nonumber\\[0.1133ex]
& \; + N \left( -\frac{3 \pi^4}{16} - 6 +
\frac{93}{2} \, \zeta(5) \right)
+ \frac{7 \pi^4}{32} \, N^2 \,.
\end{align}
Again, it is somewhat surprising that the 
leading term for large $N$ comes from the 
correction factor $\calF_\calZ$.

%
%
\subsection{Four--Point Correlation Function}
\label{sec38}

We are interested here in understanding the 
imaginary part of the four-point correlation 
function to relative order $g$. To this end, it is 
first of all necessary to remember that we only 
need to consider the first three terms
$\calK_{1,2,3}$ on the right-hand side of
Eq.~\eqref{ImW4},
because the remaining terms are of relative order $g^2$.
The dominant term, for small $g$,
is given by $\calK_1$.
We write, in analogy to Eq.~\eqref{defGab},
$\calG_{\alpha \beta \gamma \delta}$
as the sum of two terms,
the first of which is dominating,
\begin{multline}
\label{defGabcd}
\calG_{\alpha \beta \gamma \delta}(t_1, t_2, t_3, t_4) = 
\Im \calW^{(4)}_{1, \alpha \beta \gamma \delta}(t_1, t_2, t_3, t_4) 
\\
=
[ \calG_{\alpha\beta\gamma\delta}(t_1, t_2, t_3, t_4) ]_1 + 
[ \calG_{\alpha\beta\gamma\delta}(t_1, t_2, t_3, t_4) ]_2 
\\
\approx [ \calG_{\alpha\beta\gamma\delta}(t_1, t_2, t_3, t_4) ]_1 \,.
\end{multline}
The leading term is
\begin{equation}
\label{GGGGG1}
[ \calG_{\alpha\beta\gamma\delta}(t_1, t_2, t_3, t_4) ]_1 =
\frac{ \Im \calZ_{1, \alpha \beta \gamma \delta}^{(4)}(t_1,t_2,t_3,t_4) }%
{ {\calZ}_0(\undzero) } \,.
\end{equation}
The additional perturbative term, which cancels 
a few divergences, is
\begin{multline}
\label{GGGGG2}
[ \calG_{\alpha\beta\gamma\delta}(t_1, t_2, t_3, t_4) ]_2 =
- \frac{\calZ_{0, \alpha \beta}^{(2)}(t_1,t_2) 
\Im \calZ_{1, \gamma \delta}^{(2)}(t_3,t_4) }{ {\calZ}_0^2(\undzero) }
\\
- \frac{\calZ_{0,\alpha\gamma}^{(2)}(t_1,t_3) 
\Im \calZ_{1,\beta \delta}^{(2)}(t_2,t_4) }{ {\calZ}_0^2(\undzero) }
\\
- \frac{\calZ_{0,\alpha\delta}^{(2)}(t_1,t_4) 
\Im \calZ_{1, \beta\gamma}^{(2)}(t_2,t_3) }{ {\calZ}_0^2(\undzero) }
\\
- \frac{ \calZ_{0,\gamma\delta}^{(2)}(t_3,t_4) 
\Im \calZ_{1,\alpha\beta}^{(2)}(t_1,t_2) }{ {\calZ}_0^2(\undzero) }
\\
- \frac{ \calZ_{0,\beta\delta}^{(2)}(t_2,t_4) 
\Im \calZ_{1,\alpha\gamma}^{(2)}(t_1,t_4) }{ {\calZ}_0^2(\undzero) }
\\
- \frac{ \calZ_{0,\beta\gamma}^{(2)}(t_2,t_3) 
\Im \calZ_{1,\alpha \delta}^{(2)}(t_1,t_3) }{ {\calZ}_0^2(\undzero) } \,.
\end{multline}
Just as with the two-point function, 
the additional compensating perturbative terms cancel
certain divergences from diagrams that originate 
from the leading term,
in the sense of the replacement
\begin{align}
\bfDelta_\LL(t' - t_0, t_0) \to & \;
\bfDelta_\LL(t' - t_0, t_0) - \bfDelta_0(t' - t_0, t_0)  \,,
\\[0.1133ex]
\bfDelta_\TT(t' - t_0, t_0) \to & \;
\bfDelta_\TT(t' - t_0, t_0) - \bfDelta_0(t' - t_0, t_0)  \,.
\end{align}
The leading expression for the four-point 
function is easily derived, based on the 
same reasoning as was used in Eq.~\eqref{defGab},
\begin{multline}
\label{GG4leading}
\calG_{\alpha\beta\gamma\delta}(t_1, t_2, t_3, t_4) \approx 
\frac{1}{g^2} 
\langle u_\alpha \, u_\beta \, u_\gamma \, u_\delta \rangle_{S_{N-1}} \,
\\[0.1133ex]
\times \int \dd t_0  \,
\xi_\cl(t_1 - t_0) \, \xi_\cl(t_2 - t_0) \,
\xi_\cl(t_3 - t_0) \, \xi_\cl(t_4 - t_0) 
\\[0.1133ex]
= \calQ(g) \; \frac{ \delta_{\alpha\beta} \, \delta_{\gamma\delta}+
\delta_{\alpha \gamma} \, \delta_{\beta\delta} +
\delta_{\alpha \delta} \, \delta_{\gamma \beta} }{N(N+2)} \,
\frac{1}{g^2} \, J(t_1,t_2,t_3,t_4) \,,
\end{multline}
where $J(t_1,t_2,t_3,t_4)$ reads as 
\begin{multline}
\label{defJ}
J(t_1,t_2,t_3,t_4) = \int \dd t_0  \,
\xi_\cl(t_1 - t_0) \, \xi_\cl(t_2 - t_0) \,
\\[0.1133ex]
\times \xi_\cl(t_3 - t_0) \, \xi_\cl(t_4 - t_0)   
\\[0.1133ex]
= -\frac{8 (t_1 - t_4)}{\sinh(t_1 - t_2) \, \sinh(t_1 - t_3) \, \sinh(t_1 - t_4)}
\\[0.1133ex]
-\frac{8 (t_2 - t_4)}{\sinh(t_2 - t_1) \, \sinh(t_2 - t_3) \, \sinh(t_2 - t_4)}
\\[0.1133ex]
-\frac{8 (t_3 - t_4)}{\sinh(t_3 - t_1) \, \sinh(t_3 - t_2) \, 
\sinh(t_3 - t_4)} \,.
\end{multline}
This formula might seem ``asymmetric'' as the time coordinate
$t_4$ has been singled out. However, a closer 
inspection shows that the formula actually is 
symmetric with respect to a cyclic permutation 
of the time coordinates $t_i$ (with $i=1,2,3,4$).

According to Eqs.~\eqref{defSCALARG1} and~\eqref{defSCALARG2},
we can define, for the two-point function,
a ``scalar'' (with respect to the internal symmetry group)
quantity $\calG$, which is obtained from 
$\calG_{\alpha\beta}$, via division by the 
factor $\delta_{\alpha\beta}/N$. 
The same is true for the four-point function, 
where we first note that 
$\calG_{\alpha\beta\gamma\delta}$
can be written as 
\begin{multline}
\label{G4angular}
\calG_{\alpha\beta\gamma\delta}(t_1, t_2, t_3, t_4) =
\frac{ \delta_{\alpha\beta} \, \delta_{\gamma\delta}+
\delta_{\alpha \gamma} \, \delta_{\beta\delta} +
\delta_{\alpha \delta} \, \delta_{\gamma \beta} }{N(N+2)} 
\\[0.1133ex]
\times {\bfG}(t_1, t_2, t_3, t_4) \,.
\end{multline}
In leading order, one has
\begin{equation}
\label{GGJJ}
{\bfG}(t_1, t_2, t_3, t_4) \approx
\calQ(g) \, \frac{1}{g^2} \, J(t_1,t_2,t_3,t_4) \,.
\end{equation}
In turn,
${\bfG}(t_1, t_2, t_3, t_4)$ can be extracted 
from $\calG_{\alpha\beta\gamma\delta}(t_1, t_2, t_3, t_4) $ as
\begin{equation}
\label{GG4extract}
\delta_{\alpha\beta} \; \delta_{\gamma\delta} \;
\calG_{\alpha\beta\gamma\delta}(t_1, t_2, t_3, t_4) =
{\bf G}(t_1, t_2, t_3, t_4) \,.
\end{equation}
Now, one can show that 
${\bf G}(t_1, t_2, t_3, t_4)$ can be written 
as a function of the differences of the 
time coordinates only,
\begin{equation}
\label{G4trans}
{\bf G}(t_1, t_2, t_3, t_4) =
{\bf G}(t_1 - t_4, t_2 - t_4, t_3 - t_4) \,.
\end{equation}
For the leading term, given in Eq.~\eqref{defJ},
this relationship can be checked by inspection.
Let us investigate the Fourier transform
\begin{multline}
{\bf G}(p_1, p_2, p_3, p_4) =
\int \dd t_1 \, \int \dd t_2 \, \int \dd t_3 \, \int \dd t_4 
\\
\times \ee^{ -\ii ( p_1 t_1 + p_2 t_2 + p_3 t_3 + p_4 t_4 )} \,
{\bf G}(t_1 - t_4, t_2 - t_4, t_3 - t_4) \,.
\end{multline}
By a suitable change of variable,
one can show that
\begin{multline}
{\bf G}(p_1, p_2, p_3, p_4) =
2 \pi \, \delta( p_1 + p_2 + p_3 + p_4) \,
{\bf G}(p_1, p_2, p_3) 
\\[0.1133ex]
= 2 \pi \, \delta( p_1 + p_2 + p_3 + p_4) \,
\int \dd t'  \int \dd t'' \, \int \dd t''' \,
\\[0.1133ex]
\times \ee^{ -\ii ( p_1 t' + p_2 t'' + p_3 t''') } \,
{\bf G}(t', t'', t''') \,.
\end{multline}
Our task will be focused on 
\begin{multline}
\label{G4ZERODEF}
{\bf G}(p_1 = 0, p_2 = 0, p_3 = 0)  =
{\bf G}(p_{i=1,2,3} = 0) 
\\
= \int \dd t'  \int \dd t'' \, \int \dd t''' \,
{\bf G}(t', t'', t''') \,.
\end{multline}
Note that the integral $J$ defined in Eq.~\eqref{defJ} can 
be written as
\begin{subequations}
\begin{align}
\label{J4trans}
J(t_1, t_2, t_3, t_4) = & \;
J(t_1 - t_4, t_2 - t_4, t_3 - t_4) \,,
\\[0.1133ex]
\label{defJ3}
J(t', t'', t''') = & \;
-\frac{8 t'}{\sinh(t' - t'') \sinh(t' - t''') \sinh(t')}
\nonumber\\[0.1133ex]
& \; -\frac{8 t''}{\sinh(t'' - t') \sinh(t'' - t''') \sinh(t'')}
\nonumber\\[0.1133ex]
& \;
-\frac{8 t'''}{\sinh(t''' - t') \sinh(t''' - t'') \sinh(t''')},
\end{align}
\end{subequations}
so that, in leading order,
\begin{align}
\frac{{\bf G}(p_{i=1,2,3} = 0)}{\calQ(g)} \approx & \;
\frac{1}{g^2} 
\int \dd t' \int \dd t'' \int \dd t''' J(t', t'', t''') 
\nonumber\\[0.1133ex]
=& \; \frac{H_1^4}{g^2} = 
\frac{4 \pi^4}{g^2} \,.
\end{align}
It is clear that we have the same structure of corrections
as for the two-point function.
One can conveniently express the correction
as the sum of seven terms, which contribute up to 
relative order $g$,
\begin{equation}
\label{GGp_expr}
\frac{{\bf G}(p_{i=1,2,3} = 0)}{\calQ(g)} = 
\sum_{i=1}^7 \calT_i \,,
\end{equation}
where the $\calT_i$ will be defined in the following.
We have the first correction $\calT_1$ ``for free'',
because it is just the multiplicative 
correction $\calF_Z$, multiplying the leading-order
instanton result,
\begin{equation}
\calT_1 = \frac{4 \pi^4}{g^2} \, \calF_\calZ
= \frac{4 \pi^4}{g^2} +
\frac{\pi^4}{g} \left( \frac{5}{6} + \frac{9}{4} N + \frac{7}{8} N^2 \right) \,.
\end{equation}
For the Green function in coordinate space,
the replacement of two classical fields in the 
leading term
$\Im \calZ_{1, \alpha \beta \gamma \delta}^{(4)}(t_1,t_2,t_3,t_4)$ 
by two fluctuations leads to 
two terms $[ \calT_2 ]_1$ and $[ \calT_3 ]_1$,
where the subscript $[ \ldots ]_1$ is motivated
by Eq.~\eqref{GGGGG1}.
After suitable variables changes, 
their contribution to the four-point function at zero momentum
can be written as
\begin{align}
[ \calT_2 ]_1 = & \;
-\frac{6}{g} \int \dd t_0 \int \dd t' \int \dd t'' \int \dd t'''
\nonumber\\[0.1133ex]
& \; \times \xi_\cl(t_0) \, \xi_\cl(t''') \, \Delta_\LL(t', t'') \,,
\\[0.1133ex]
[ \calT_3 ]_1 = & \;
-\frac{2(N-1)}{g} \int \dd t_0 \int \dd t' \int \dd t'' \int \dd t'''
\nonumber\\[0.1133ex]
& \; \times \xi_\cl(t_0) \, \xi_\cl(t''') \, \Delta_\TT(t', t'') \,.
\end{align}
In order to evaluate the compensating perturbative
terms $[ \calT_2 ]_2$ and $[ \calT_3 ]_2$
from Eq.~\eqref{GGGGG2},
one replaces [see Eq.~\eqref{ZGalphabeta}]
\begin{equation}
\frac{\calZ_{0, \alpha \beta}^{(2)}(t_1,t_2) }{ {\calZ}_0(\undzero) }
\to \frac{\delta_{\alpha\beta}}{N} \,
[ N \, \bfDelta_0(t_1, t_2) ] \,.
\end{equation}
(There is no integration over a collective coordinate here,
as we are analyzing the perturbative propagator.)
For the term involving the instanton saddle point, we have
\begin{align}
\frac{\Im \calZ_{1, \gamma \delta}^{(2)}(t_3,t_4) }%
{ {\calZ}_0(\undzero) } \to & \;
\left(  -\frac{1}{g} \right)
\frac{\delta_{\gamma\delta}}{N} \, \calQ(g) 
\nonumber\\[0.1133ex]
& \; \times \int \dd t_0 \,
\xi_\cl(t_3 - t_0) \, \xi_\cl(t_4 - t_0) .
\end{align}
After suitable variable changes, 
one arrives at the following ``counter-''term from the
expression in Eq.~\eqref{GGGGG2},
\begin{align}
[ \calT_2 ]_2 =& \; \frac{6}{g} \int \dd t_0 \int \dd t' \int \dd t'' \int \dd t'''
\nonumber\\[0.1133ex]
& \; \times \xi_\cl(t_0) \, \xi_\cl(t''') \, \Delta_0(t', t'') \,,
\\[0.1133ex]
[ \calT_3 ]_2 =& \; \frac{2(N-1)}{g} \int \dd t_0 \int \dd t' \int \dd t'' \int \dd t'''
\nonumber\\[0.1133ex]
& \; \times \xi_\cl(t_0) \, \xi_\cl(t''') \, \Delta_0(t', t'') \,.
\end{align}
Hence, $\calT_2$ and $\calT_3$ can be expressed as
\begin{align}
\label{FFF2}
\calT_2 = & \; 
[ \calT_2 ]_1 + [ \calT_2 ]_2
= -\frac{6}{g} \, H_1^2 \, I_1 = \frac{48 \, \pi^2}{g} \,,
\\[0.1133ex]
\label{FFF3}
\calT_3 = & \; 
[ \calT_3 ]_1 + [ \calT_3 ]_2
= \frac{N-1}{g} \left( \pi^4 - 4 \pi^2 + 14 \pi^2 \zeta(3) \right) \,.
\end{align}
Note that $\calT_2$ and $\calT_3$ are analogous to the 
terms $G_2$ and $G_3$, incurred for the two-point function.
The integrals $H_1$, $I_1$ and $I_2$ can be found in 
Appendix~\ref{appendixb}.

Now we must treat the analogues of the 
terms $G_{4,5,6,7}$, generated by the 
Jacobian factor $\calF_\calJ$, originally derived
for the two-point function.
These terms are generated by the replacement of 
one classical field configuration by a fluctuation,
combined, via the Wick theorem, with 
a contraction with a second fluctuation in the 
action$+$Jacobian factor $\calF_\calJ$.
For the two-point function, we have two possibilities
to choose one field out of two, for the cases where
only one field is replaced by a fluctuation.
For the four-point function, we have four such possibilities,
so, we could tentatively conjecture that 
the corrections $\calT_4$, $\calT_5$, $\calT_6$ and $\calT_7$ 
receive a relative factor two, 
as compared to the two-point Green function.
This will turn out to be a good guess,
but it needs to be verified by an explicit calculation.

We now consider the term
\begin{equation}
\sqrt{-g} \, 
\int \dd t \und \xi_{\rm cl}(t) \cdot \und\chi(t) \;\; \und \chi^2(t)
\end{equation}
from Eq.~\eqref{defFJ}. One obtains two corrections 
$\calT_4$ and $\calT_5$ to 
${\bf G}(p_{i=1,2,3} = 0)/\calQ(g)$,
according to Eq.~\eqref{GGp_expr},
\begin{align}
\calT_4 = & \; - \frac{12}{g} \int \dd t_0 \int \dd t' \int \dd t'' \int \dd t'''
\int \dd t \, \xi_\cl(t_0) \, \xi_\cl(t''') \, 
\nonumber\\[0.1133ex]
& \; 
\times \xi_\cl(t') \, \xi_\cl(t) \, \Delta_\LL(t, t) \, \Delta_\LL(t, t'') \,,
\nonumber\\[0.1133ex]
=& \; -\frac{12}{g} H_1^3 \, J_3 = -\frac{6 \pi^4}{g} \,,
\\[0.1133ex]
\calT_5 = & \;
-\frac{4 (N-1)}{g} \int \dd t_0 \int \dd t' \int \dd t'' \int \dd t'''
\int \dd t \,
\nonumber\\[0.1133ex]
& \; \times \xi_\cl(t_0) \, \xi_\cl(t''') \, 
\xi_\cl(t') \, \xi_\cl(t) \, 
\Delta_\TT(t, t) \, \Delta_\LL(t, t'') 
\nonumber\\[0.1133ex]
=& \; -\frac{4 ( N - 1 ) }{g} H_1^3 \, J_4
= -\frac{\pi^4}{g} \, \left( N - 1 \right) \,.
\end{align}
From the term
\begin{equation}
\sqrt{-g} \, 
\left[ - \frac{3}{4} \int \dd t \, \und \chi(t) \cdot 
\ddot{\und \xi}_{\rm cl}(t) \right] 
\end{equation}
in Eq.~\eqref{defFJ}, one has the correction
\begin{align}
\calT_6 = & \;
\frac{3}{g} \int \dd t_0 \int \dd t' \int \dd t'' \int \dd t'''
\int \dd t \,
\nonumber\\[0.1133ex]
& \; \times \xi_\cl(t_0) \, \xi_\cl(t''') \, 
\xi_\cl(t') \, \ddot \xi_\cl(t) \, 
\Delta_\LL(t, t'') 
\nonumber\\[0.1133ex]
=& \; \frac{3}{g} H_1^3 \, J_2
= \frac{6 \pi^4}{g} \,.
\end{align}
The last correction is from the term
\begin{equation}
\sqrt{-g} \, 
\frac{N-1}{4} 
\int\dd t\,{\underline \chi}(t)\cdot {\underline \xi}_{\rm cl}(t)
\end{equation}
in Eq.~\eqref{defFJ} and reads as
\begin{align}
\calT_7 = & \;
-\frac{N-1}{g} \int \dd t_0 \int \dd t' \int \dd t'' \int \dd t'''
\nonumber\\[0.1133ex]
& \; \times \int \dd t \,
\xi_\cl(t_0) \, \xi_\cl(t''') \,
\xi_\cl(t') \, \xi_\cl(t) \,
\Delta_\LL(t, t'')
\nonumber\\[0.1133ex]
=& \; \frac{N-1}{g} H_1^3 \, J_1 = 0 \,.
\end{align}
The end result is given as follows,
\begin{multline}
\label{GGGGGRES}
\frac{{\bf G}(p_{i=1,2,3} = 0)}{\calQ(g)} = 
\sum_{i=1}^7 \calT_i 
= \frac{4 \pi^4}{g^2} 
\nonumber\\[0.1133ex]
+ \frac{\pi^2}{g} \,
\left[ 52 + \frac{5 \pi^2}{6} - 14 \zeta(3) 
\right.
\nonumber\\[0.1133ex]
\left. + N \, \left( \frac{9 \pi^2}{4} - 4 + 14 \zeta(3) \right) 
+ \frac{7 \pi^2}{8} N^2 \right] \,,
\end{multline}
involves, again, a couple of Riemann zeta functions.
Finally, we recall the Fourier transform
of Eq.~\eqref{G4angular} in the form
\begin{align}
\frac{\calG_{\alpha\beta\gamma\delta}(p_{i=1,2,3} = 0)}{\calQ(g)}
=& \; \frac{ \delta_{\alpha\beta} \, \delta_{\gamma\delta}+
\delta_{\alpha \gamma} \, \delta_{\beta\delta} +
\delta_{\alpha \delta} \, \delta_{\gamma \beta} }{N(N+2)} 
\nonumber\\[0.1133ex]
& \; \times \frac{{\bf G}(p_{i=1,2,3} = 0)}{\calQ(g)} \,,
\end{align}
exhibiting the angular structure within the 
internal $O(N)$ group.

%
%
\subsection{Two--Point Wigglet Insertion}
\label{sec39}

In this section we will derive the large order behavior of the two-point
correlation function with a wigglet insertion computed at zero momentum. Using
the previous results relative to the two- and four- point function, its
derivation will be straightforward.
We define the (imaginary part of the) wigglet insertion 
into the two-point Green function
$\calG_{\alpha \beta}(t_1, t_2)$, according to Eq.~\eqref{ImW},
as follows,
\begin{align}
\label{defGab2}
\calG^{(1,2)}_{\alpha \beta}(t_1, t_2) = & \;
\left.  \frac{\partial^2}{\partial m^2} 
\calG_{\alpha \beta}(m, t_1, t_2) 
\right|_{m^2 = 1} 
\nonumber\\[0.1133ex]
=& \; 
\left.  \frac{\partial^2}{\partial m^2} 
\Im \calW^{(2)}_{1, \alpha \beta}(m, t_1, t_2) 
\right|_{m^2 = 1} 
\nonumber\\[0.1133ex]
=& \; 
-\dfrac{1}{2} \int \dd s 
\Im \calW^{(4)}_{\alpha\beta\gamma\gamma}(s, s, t_1, t_2)
\nonumber\\[0.1133ex]
=& \; -\dfrac{1}{2} \int \dd s \, 
\calG_{\alpha\beta\gamma\gamma}(s, s, t_1, t_2) \,.
\end{align}
Here, $\calG_{\alpha \beta}(m, t_1, t_2)$
is the analogue of $\calG_{\alpha \beta}(t_1, t_2)$,
defined in Eq.~\eqref{defGab}, but with respect
to an action with a variable mass term,
\begin{equation}
\label{SSmqq}
\calS[m, q(t)] = \int \dd t \, 
\left( \frac12 \, \dot q^2 + \frac{m^2}{2} \, q^2 + 
\frac14 \, g \, q^4 \right) \,.
\end{equation}
In view of the result~\eqref{GG4leading}
for the leading contribution to the 
four-point function, 
we have the relation [see Eq.~\eqref{GG4leading}]
\begin{align}
\label{G12leading}
\calG^{(1,2)}_{\alpha \beta}(t_1, t_2) = & \;
-\dfrac{1}{2} \int \dd s \, 
\calG_{\alpha\beta\gamma\gamma}(s, s, t_1, t_2) 
\nonumber\\[0.1133ex]
= & \; -\frac{\calQ(g)}{ 2 } \;
\frac{ \delta_{\alpha\beta} }{N} \,
\frac{1}{g^2} \, \frac{16 (t_1 - t_2) }{ \sinh(t_1 - t_2) } \,,
\end{align}
where we have used Eq.~\eqref{defcalQ}.
One can show that
$\calG^{(1,2)}_{\alpha \beta}(t_1, t_2)$ can be written
in terms of the time coordinate differences only,
\begin{equation}
\calG^{(1,2)}_{\alpha \beta}(t_1, t_2) = 
\calG^{(1,2)}_{\alpha \beta}(t_1 - t_2) \,.
\end{equation}
In leading order, we anticipate the result
\begin{align}
\label{G12p0leading}
\int \dd(t_1-t_2) \; \calG^{(1,2)}_{\alpha \beta}(t_1 - t_2) 
\approx & \;
-\frac{\calQ(g)}{ 2 } \; \frac{ \delta_{\alpha\beta} }{N} \,
\frac{H_1^2 \, H_2}{g^2} 
\nonumber\\[0.1133ex]
=& \; -\frac{\calQ(g)}{ 2 } \; \frac{ \delta_{\alpha\beta} }{N} \,
\frac{8 \pi^2}{g^2} \,,
\end{align}
where we have used, again, integrals 
from Appendix~\ref{appendixb}.
In order to analyze the two-loop corrections,
we first need to remember the angular structure.
First, it is easy to see that, from Eq.~\eqref{G4angular},
we have
\begin{equation}
\calG_{\alpha\beta\gamma\gamma}(t_1, t_2, t_3, t_4) =
\frac{ \delta_{\alpha\beta} }{ N } \, 
{\bf G}(t_1, t_2, t_3, t_4) \,.
\end{equation}
So,
\begin{align}
\label{defGab12}
\calG^{(1,2)}_{\alpha \beta}(t_1, t_2) 
=& \; -\frac{ \delta_{\alpha\beta} }{ 2 N } \, 
\int \dd s \, {\bf G}(s, s, t_1 - t_2) 
\nonumber\\[0.1133ex]
=& \; -\frac{ \delta_{\alpha\beta} }{ 2 N } \, {\bf H}(t_1, t_2) 
= -\frac{ \delta_{\alpha\beta} }{ 2 N } \, {\bf H}(t_1 - t_2) \,,
\end{align}
where we appeal to Eq.~\eqref{G4trans} and
implicitly define the function
${\bf H}(t_1, t_2) = {\bf H}(t_1 - t_2)$.
Let us investigate the Fourier transform
\begin{align}
{\bf H}(p_1, p_2) = & \;
\int \dd t_1 \, \int \dd t_2 \, 
\ee^{ -\ii ( p_1 t_1 + p_2 t_2 )} \,
\int \dd s \, 
{\bf H}(t_1, t_2) 
\nonumber\\[0.1133ex]
=& \; 2 \pi \, \delta(p_1 + p_2) \, {\bf H}(p_1) \,,
\end{align}
where
\begin{equation}
{\bf H}(p) = \int \dd t' \, \int \dd t'' \, 
\ee^{ -\ii p t'' } \, {\bf G}(t', t', t'') \,,
\end{equation}
so that
\begin{equation}
\label{zeromom_twoW}
{\bf H}(p = 0) 
= \int \dd s \, \int \dd (t_1 - t_2) \, {\bf G}(s, s, t_1, t_2) \,,
\end{equation}
where we again use Eq.~\eqref{G4trans}.
In leading order, in view of Eqs.~\eqref{GGJJ} 
and~\eqref{defJ3}, one has
\begin{align}
\frac{{\bf H}(p = 0)}{\calQ(g)} \approx & \;
\frac{1}{g^2} \int \dd t' \, \int \dd t'' \, J(t', t', t'') 
\nonumber\\[0.1133ex]
=& \; \frac{H_1^2 \, H_2}{g^2} 
= \frac{8 \pi^2}{g^2} \,,
\end{align}
confirming the leading-order 
result given in Eq.~\eqref{G12p0leading}.
One can conveniently express the correction
as the sum of seven terms, which contribute up to
relative order $g$,
\begin{equation}
\label{HHp_expr}
\frac{{\bf H}(p = 0)}{\calQ(g)} = 
\sum_{i=1}^7 \calU_i \,.
\end{equation}
We have the first correction ``for free'',
because it is just the multiplicative 
correction $\calF_Z$, multiplying the leading-order
instanton result,
\begin{equation}
\calU_1 = \frac{8 \pi^2}{g^2} \, \calF_\calZ
= \frac{8 \pi^2}{g^2} +
\frac{2 \pi^2}{g} \left( \frac{5}{6} + \frac{9}{4} N + \frac{7}{8} N^2 \right) \,.
\end{equation}
Just as we did in our analysis of the four-point function,
we now replace, in
the Green function in coordinate space,
two classical fields by fluctuations,
and obtain two terms,
\begin{align}
[ \calU_2 ]_1 = & \;
-\frac{1}{g} \int \dd t' \int \dd t'' \, \Delta_\LL(t', t'') \,
\int \dd t''' \, [ \xi_\cl(t''') ]^2 
\nonumber\\[0.1133ex]
& \; -\frac{4}{g} \int \dd t' \xi_\cl(t') \, 
\int \dd t'' \, \Delta_\LL(t', t'') \,
\int \dd t''' \, \xi_\cl(t''')
\nonumber\\[0.1133ex]
& \; -\frac{1}{g} \int \dd t' \Delta_\LL(t', t') \,
\int \dd t'' \, \xi_\cl(t'') \,
\int \dd t''' \, \xi_\cl(t''')
\,,
\\[0.1133ex]
[ \calU_3 ]_1 = & \;
-\frac{N-1}{g} \int \dd t' \int \dd t'' \, \Delta_\TT(t', t'') \,
\int \dd t''' \, [ \xi_\cl(t''') ]^2 
\nonumber\\[0.1133ex]
& \; -\frac{N-1}{g} \int \dd t' \, \Delta_\TT(t', t') 
\nonumber\\[0.1133ex]
& \; \times \int \dd t'' \, \xi_\cl(t'') \,
\int \dd t''' \, \xi_\cl(t''') \,. 
\end{align}
From the perturbative compensating term in Eq.~\eqref{GGGGG2},
we have
\begin{align}
[ \calU_2 ]_2 =& \; 
\frac{1}{g} \int \dd t' \int \dd t'' \, \Delta_0(t', t'') \,
\int \dd t''' \, [ \xi_\cl(t''') ]^2 
\nonumber\\[0.1133ex]
& \; + \frac{4}{g} \int \dd t' \xi_\cl(t') \, 
\int \dd t'' \, \Delta_0(t', t'') \,
\int \dd t''' \, \xi_\cl(t''')
\\[0.1133ex]
& \; -\frac{1}{g} \int \dd t' \Delta_0(t', t') \,
\int \dd t'' \, \xi_\cl(t'') \,
\int \dd t''' \, \xi_\cl(t''') \,,
\nonumber\\[0.1133ex]
[ \calU_3 ]_2 =& \; 
\frac{N-1}{g} \int \dd t' \int \dd t'' \, \Delta_0(t', t'') \,
\int \dd t''' \, [ \xi_\cl(t''') ]^2 
\nonumber\\[0.1133ex]
& \; + \frac{N-1}{g} \int \dd t' \, \Delta_0(t', t') \,
\nonumber\\[0.1133ex]
& \; \times \int \dd t'' \, \xi_\cl(t'') \,
\int \dd t''' \, \xi_\cl(t''') \,. 
\end{align}
With the help of the integrals
listed in Appendix~\ref{appendixb}, 
we can express $\calU_2$ and $\calU_3$ as follows,
\begin{align}
\calU_2 = & \;
-\frac{1}{g} H_2 \, I_1 
-\frac{4}{g} H_1 \, (J_1 - J_1^{(0)}) 
-\frac{1}{g} H_1^2 \, N_1 
\nonumber\\[0.1133ex]
=& \; \frac{ 48 + 31 \pi^2 }{ 3 \, g } \,,
\\[0.1133ex]
\calU_3 = & \;
-\frac{N-1}{g} H_2 \, I_2
-\frac{N-1}{g} H_1^2 \, N_2 
\nonumber\\[0.1133ex]
=& \; \frac{2 (N-1)}{g} 
\left[ \pi^2 - 2 + 7 \, \zeta(3) \right] \,.
\end{align}
We now consider the term
\begin{equation}
\sqrt{-g} \, \left[ \int \dd t \und \xi_{\rm cl}(t) \cdot 
\und\chi(t) \;\; \und \chi^2(t) \right] 
\end{equation}
in the Jacobian Eq.~\eqref{defFJ}.
After the application of the Wick theorem and
suitable variable changes, 
one arrives at a contribution to
${\bf H}(p = 0)/\calQ(g)$,
consisting of the sum of two terms, $\calU_4$ and $\calU_5$,
\begin{align}
\calU_4 = & \;
-\frac{6}{g} \int \dd t \, \xi_\cl(t) \, \Delta_\LL(t, t) \, 
\nonumber\\[0.1133ex]
& \; \times \int \dd t' \Delta_\LL(t, t') \int \dd t'' [\xi_\cl(t'')]
\int \dd t''' [\xi_\cl(t''')]^2 
\nonumber\\[0.1133ex]
& \; -\frac{6}{g} \int \dd t \, \xi_\cl(t) \, \Delta_\LL(t, t) \, 
\int \dd t' \xi_\cl(t') \, \Delta_\LL(t, t') \,
\nonumber\\[0.1133ex]
& \; \times \left[ \int \dd t'' \xi_\cl(t'') \right]^2
\nonumber\\[0.1133ex]
=& \; -\frac{6}{g} H_1 \, H_2 \, J_3 
- \frac{6}{g} H_1^2 \, M_3 
= -\frac{19 \pi^2}{3 \, g} \,,
\\[0.1133ex]
\calU_5 = & \;
-\frac{2(N-1)}{g} \int \dd t \, \xi_\cl(t) \, \Delta_\TT(t, t) \, 
\int \dd t' \Delta_\LL(t, t') 
\nonumber\\[0.1133ex]
& \; \times \int \dd t'' [\xi_\cl(t'')]
\int \dd t''' [\xi_\cl(t''')]^2 
\nonumber\\[0.1133ex]
& \; -\frac{2(N-1)}{g} \int \dd t \, \xi_\cl(t) \Delta_\TT(t, t) 
\int \dd t' \, \xi_\cl(t') \Delta_\LL(t, t') 
\nonumber\\[0.1133ex]
& \; \times \left[ \int \dd t'' \xi_\cl(t'') \right]^2
\nonumber\\[0.1133ex]
=& \; -\frac{2 ( N - 1 ) }{g} H_1 \, H_2 \, J_4
- \frac{2 ( N - 1 ) }{g} H_1^2 \, M_4 = 0 \,.
\end{align}
From the term
\begin{equation}
\sqrt{-g} \, 
\left[ - \frac{3}{4} \int \dd t \, \und \chi(t) \cdot 
\ddot{\und \xi}_{\rm cl}(t) \right] 
\end{equation}
in the Jacobian given in Eq.~\eqref{defFJ}, one has
\begin{align}
\calU_6 = & \;
\frac{3}{2 g} \int \dd t \xi_\cl(t) \, 
\int \dd t' \Delta_\LL(t, t') \,
\int \dd t'' \, \xi_\cl(t'') \, 
\nonumber\\[0.1133ex]
& \; \times \int \dd t''' \, [\xi_\cl(t''')]^2 \, 
\nonumber\\[0.1133ex]
& \; + \frac{3}{2 g} \int \dd t \xi_\cl(t) \, 
\int \dd t' \xi_\cl(t') \, \Delta_\LL(t, t') \,
\nonumber\\[0.1133ex]
& \; \times \left[ \int \dd t'' \, \xi_\cl(t'') \right]^2 \, 
\nonumber\\[0.1133ex]
=& \; 
\frac{3}{2 g} H_1 \, H_2 \, J_2
+ \frac{3}{2 g} H_1^2 \, M_2
= \frac{9 \pi^2}{g} \,.
\end{align}
The last correction is from the term
\begin{equation}
\sqrt{-g} \, 
\frac{N-1}{4} 
\int\dd t\,{\underline \chi}(t)\cdot {\underline \xi}_{\rm cl}(t)
\end{equation}
in Eq.~\eqref{defFJ} and leads to the correction
\begin{align}
\calU_7 = & \;
-\frac{N-1}{2 g} 
\int \dd t \xi_\cl(t) \,
\int \dd t' \, \Delta_\LL(t, t') 
\nonumber\\[0.1133ex]
& \; \times \int \dd t'' \xi_\cl(t'') \,
\int \dd t''' \, [\xi_\cl(t''')]^2 \,
\nonumber\\[0.1133ex]
& \; -\frac{N-1}{2 g} 
\int \dd t \xi_\cl(t) \,
\int \dd t' \, \xi_\cl(t') \, \Delta_\LL(t, t') \,
\nonumber\\[0.1133ex]
& \; \times \left[ \int \dd t'' \xi_\cl(t'') \right]^2 \,
\nonumber\\[0.1133ex]
=& \; 
-\frac{N-1}{2 g} H_1 \, H_2 \, J_1 
-\frac{N-1}{2 g} H_1^2 \, M_1 =
\frac{\pi^2 (N-1)}{g} \,.
\end{align}
The final result [see Eq.~\eqref{GGp_expr}]
for order-$g$ corrections to the 
imaginary part of the wigglet insertion into the 
two-point function reads as follows,
\begin{multline}
\label{G12RES}
\frac{{\bf H}(p = 0)}{\calQ(g)} =
\sum_{i=1}^7 \calU_i = 
\frac{8 \pi^2}{g^2} 
+ \frac{1}{g} \,
\left[ 20 + \frac{35 \pi^2}{3} - 14 \, \zeta(3) 
\right.
\\[0.1133ex]
\left. +
N \, \left( \frac{15 \pi^2}{2} - 4 + 14 \zeta(3) \right) 
+ \frac{7 \pi^2}{4} N^2 \right] \,.
\end{multline}
It has the same structure as the result for 
the four-point function, given in 
Eq.~\eqref{GGGGGRES}.

%
%
\section{Conclusions}
\label{sec4}

%
%
\subsection{Large--Order Behavior: A Summary}
\label{sec41}

In this article, we have concentrated on the one-dimensional 
field theory, with an internal $O(N)$ symmetry group,
in the normalization
[see Eq.~\eqref{SON}]
\begin{align}
{\calS}[\und q(t)] =&  \int \dd t \, \left[ 
\frac12 \, \left( \frac{\partial \und q(t)}{\partial t} \right)^2 +
\frac12 \, \und q^2(t) + \frac{g}{4} \, \und q^4(t) \right] \,,
\end{align}
where
$\und q(t) = \{ q_1(t), \dots, q_N(t) \} 
= \sum_{\alpha = 1}^N q_\alpha(t) \, \und e_\alpha$.
The start time $t_0$ of the instanton,
and the coordinate $\tau_i$ with $i = 1,\dots, N-1$,
are the collective coordinates of the problem.
Furthermore, we have analyzed the functional 
determinant of the transformation 
of the path integral, into integrals over the 
collective coordinates and path integrals 
over paths orthogonal to the zero modes,
in Sec.~\ref{sec23}, with the result 
for the Jacobian given in Eq.~\eqref{JACRES}.

Based on the results given in Sec.~\ref{sec3} and the 
dispersion relations studied in Sec.~\ref{sec12},
we are now in the position to write down the 
large-order behavior of various perturbative
expansions, of the partition function and 
various correlation functions
[in the sense of Eqs.~\eqref{ImGM} and~\eqref{genexp}]. 
We have from Eq.~\eqref{Z1corr},
for the imaginary part of the ground-state energy of 
the $O(N)$ oscillator,
\begin{align}
\frac{{\rm Im} \, E_0(g)}{\calQ(g)} =& \;
\lim_{\beta \to \infty} 
\left( -\frac{1}{\calQ(g) \; \beta} \;
\frac{{\rm Im} \; \calZ_1(\beta)}{\calZ_{0}(\beta)} \right)
\nonumber\\[0.1133ex]
=& \; - \left[ 1 + 
g \, \left( \frac{7}{32} N^2 + \frac{9}{16} N + \frac{5}{24} \right) 
\right] \,.
\end{align}
Here, according to Eq.~\eqref{defcalQ}, one has
\begin{equation}
\calQ(g) = \frac{1}{\Gamma(N/2)} \, 
\left( - \frac{8}{g} \right)^{N/2} \,
\exp\left( \frac{4}{3 \, g} \right) \,.
\end{equation}
Based on the formalism outlined in 
Sec.~\ref{sec12}, this can be converted 
to the asymptotics of the 
perturbative coefficients of the ground-state 
energy in large orders [see Eq.~\eqref{Z1corr}].
One can identify $E_0(g) \equiv \sfG^{(0)}(g)$ 
with the ground-state energy, and 
study the perturbative expansion 
$\sfG^{(0)}(g) = \sum_K \sfG^{(0)}_K \, g^K$.
Within the conventions outlined in
Eq.~\eqref{genexp}, one has $n=0$, $D=1$ and 
\begin{subequations}
\begin{align}
c^{(0)}(N,1) =& \; -\frac{8^{N/2}}{\Gamma(N/2)},\\
d^{(0)}(N,1) =& \; 
\frac{7}{32} N^2 + \frac{9}{16} N + \frac{5}{24} \,.
\end{align}
\end{subequations}
The two-point function has the 
angular structure [see Eqs.~\eqref{defSCALARG1} 
and~\eqref{GDIFF}]
\begin{equation}
{\rm Im} \, \sfG^{(2)}_{\alpha\beta}(t_1 - t_2) = 
\calG_{\alpha\beta}(t_1,t_2) = 
\frac{\delta_{\alpha\beta}}{N} \calG(t_1 - t_2).
\end{equation}
We recall that $\calG_{\alpha\beta}(t_1, t_2)$,
as defined in Eq.~\eqref{defGab}, is  
the imaginary part of the two-point function 
$\sfG_{\alpha\beta}(t_1 - t_2)$, 
in the sense in Sec.~\ref{sec12}.  The result for
$\calG(p = 0) = \int \dd t' \calG(t')$ of the
``scalar'' two-point function 
at $p=0$ has been recorded in Eq.~\eqref{GP0RES},
\begin{align}
\frac{\calG(p = 0)}{ \calQ(g) } =& \; - \frac{2 \pi^2}{g} \, \left[ 1 
+ g \, \left\{
\frac{5}{24} + \frac{5}{2 \pi^2} - \frac{7 \zeta(3)}{4 \pi^2} 
\right.
\right.
\nonumber\\[0.1133ex]
& \; \left.  \left.
+ \left( \frac{9}{16} - \frac{1}{2 \pi^2} + \frac{7 \zeta(3)}{4 \pi^2} \right) \, N
+ \frac{7}{32}  \, N^2 \right\} \right] \,.
\end{align}
For the imaginary part $\calG(p = 0)$,
this translates into the following 
coefficients in the large-order asymptotics,
according to Eq.~\eqref{genexp},
\begin{subequations}
\begin{align}
c^{(2)}(N,1) =& \;
2\pi^2\frac{8^{N/2}}{\Gamma(N/2)},\\
d^{(2)}(N,1) =& \; 
\frac{5}{24} + \frac{5}{2 \pi^2} - \frac{7 \zeta(3)}{4 \pi^2} 
\nonumber\\[0.1133ex]
& \; + \left( \frac{9}{16} - \frac{1}{2 \pi^2} + \frac{7 \zeta(3)}{4 \pi^2} \right) N
+ \frac{7}{32}  \, N^2 \,.
\end{align}
\end{subequations}
Of course, we have $n=2$ in the sense of
Eq.~\eqref{genexp} for the two-point function 
at zero momentum transfer as well as 
for its derivative, which 
gives rise to the following 
imaginary part, according to Eq.~\eqref{GPP0},
\begin{multline}
\frac{\left. \dfrac{\partial^2}{\partial p^2} \calG \right|_{p = 0}}{\calQ(g)} = 
\frac{\pi^4}{g} 
\left[ 1 + g \left\{
\frac{5}{24} + \frac{4}{\pi^4} 
- \frac{21 \zeta(3)}{\pi^4} 
\right.
\right.
\\[0.1133ex]
\left.  \left.
- \frac{93 \zeta(5)}{2 \pi^4} 
+ \left( -\frac{3}{16} - \frac{6}{\pi^4} + \frac{93 \zeta(5)}{2 \pi^4} \right) N
+ \frac{7 N^2}{32} \right\} \right].
\end{multline}
The leading and subleading large-order asymptotics 
for the perturbative coefficients of 
$\left. \frac{\partial^2}{\partial p^2} \calG \right|_{p = 0}$
are given as follows [see Eq.~\eqref{genexp}],
\begin{subequations}
\begin{align}
c^{(\partial p)}(N,1) =& \;
-\pi^4\frac{8^{N/2}}{\Gamma(N/2)},\\
d^{(\partial p)}(N,1) =& \;
\frac{5}{24} + \frac{4}{\pi^4} 
- \frac{21 \, \zeta(3)}{\pi^4} 
- \frac{93 \, \zeta(5)}{2 \pi^4} 
\\[0.1133ex]
& \;
+ \left( -\frac{3}{16} - \frac{6}{\pi^4} + \frac{93 \zeta(5)}{2 \pi^4} \right) \, N
+ \frac{7}{32}  \, N^2 \,.
\nonumber
\end{align}
\end{subequations}
For the imaginary part of the four-point function,
according to Eq.~\eqref{G4angular}, we have
\begin{multline}
\calG_{\alpha\beta\gamma\delta}(t_1, t_2, t_3, t_4) =
\frac{ \delta_{\alpha\beta} \, \delta_{\gamma\delta}+
\delta_{\alpha \gamma} \, \delta_{\beta\delta} +
\delta_{\alpha \delta} \, \delta_{\gamma \beta} }{N(N+2)} 
\\
\times
{\bf G}(t_1 - t_4, t_2 - t_4, t_3 - t_4) \,.
\end{multline}
One defines according to Eq.~\eqref{G4ZERODEF},
\begin{multline}
{\bf G}(p_1 = 0, p_2 = 0, p_3 = 0) =
{\bf G}(p_{i=1,2,3} = 0) 
\\[0.1133ex]
= \int \dd t'  \int \dd t'' \, \int \dd t''' \,
{\bf G}(t', t'', t''') \,.
\end{multline}
The result, to relative order $g$, is 
given by [according to Eq.~\eqref{GGGGGRES}],
\begin{multline}
\frac{{\bf G}(p_{i=1,2,3} = 0)}{\calQ(g)} = 
\frac{4 \pi^4}{g^2} 
\left[ 1 + g \, \left\{
\frac{5}{24} + \frac{13}{\pi^2} 
- \frac{7 \, \zeta(3)}{2 \pi^2} 
\right.
\right.
\\[0.1133ex]
\left.  \left.
+ \left( \frac{9}{16} - \frac{1}{\pi^2} + \frac{7 \zeta(3)}{2 \pi^2} \right) \, N
+ \frac{7 N^2}{32} \right\} \right].
\end{multline}
The perturbative expansion of the 
imaginary part ${\bf G}(p_{i=1,2,3} = 0)$
of the four-point function therefore has the 
following asymptotics of the
perturbative coefficients [see Eq.~\eqref{genexp}
with $n=4$]
\begin{subequations}
\begin{align}
c^{(4)}(N,1) =& \;
4\pi^4\frac{8^{N/2}}{\Gamma(N/2)},\\
d^{(4)}(N,1) =& \; 
\frac{5}{24} + \frac{13}{\pi^2} 
- \frac{7 \, \zeta(3)}{2 \pi^2} 
\\[0.1133ex]
& \; + \left( \frac{9}{16} - \frac{1}{\pi^2} + \frac{7 \zeta(3)}{2 \pi^2} \right) \, N
+ \frac{7}{32}  \, N^2 \,.
\nonumber
\end{align}
\end{subequations}
Finally, for the imaginary part of the 
two-point function with a wigglet insertion, 
computed at zero momentum, we have the following expression, 
according to Eq.~\eqref{defGab12},
\begin{equation}
\calG^{(1,2)}_{\alpha \beta}(t_1, t_2) =
-\frac{ \delta_{\alpha\beta} }{2 N} \,
{\bf H}(t_1 - t_2) \,.
\end{equation}
The quantity of interest is [see Eq.~\eqref{zeromom_twoW}]
\begin{equation}
{\bf H}(p = 0) = \int \dd t' \, {\bf H}(t')  \,,
\end{equation}
for which we obtain the result [see Eq.~\eqref{G12RES}]
\begin{multline}
\frac{{\bf H}(p = 0)}{\calQ(g)} =
\frac{8 \pi^2}{g^2} \left[ 1 + g \left\{
\frac{35}{24} + \frac{5}{2 \pi^2} 
- \frac{7 \, \zeta(3)}{4 \pi^2} 
\right.
\right.
\\[0.1133ex]
\left.
\left.
+ \left( \frac{15}{16} - \frac{1}{2 \pi^2} + \frac{7 \zeta(3)}{4 \pi^2} \right) \, N
+ \frac{7}{32}  \, N^2 \right\} \right] \,.
\end{multline}
The large-order asymptotics of the perturbative coefficients
for ${\bf H}(t_1 - t_2)$ are as follows,
\begin{subequations}
\begin{align}
c^{(1,2)}(N, 1) =& \;
8\pi^2\frac{8^{N/2}}{\Gamma(N/2)}, \\
d^{(1,2)}(N, 1) = & \;
\frac{35}{24} + \frac{5}{2 \pi^2} 
- \frac{7 \, \zeta(3)}{4 \pi^2} 
\nonumber\\[0.1133ex]
& \;
+ \left( \frac{15}{16} - \frac{1}{2 \pi^2} + \frac{7 \zeta(3)}{4 \pi^2} \right) \, N
+ \frac{7}{32}  \, N^2 \,.
\end{align}
\end{subequations}
We note that the large-order asymptotics have the parameter
$n=4$ in the conventions delineated in Eq.~\eqref{genexp},
because of the additional two fields that have to be inserted 
in view of the mass derivative.

%
%
\subsection{Interpretation of the Results}

In this article, we have laid the groundwork for the 
accurate systematic analysis of the 
large-order behavior of perturbation theory 
for the correlation functions
in the $N$-vector model.
Our paradigm is that once the number of loops in a Feynman
diagram becomes very large, the large-order behavior
of the $N$-vector model is determined by classical 
field configurations (instantons), 
which determine the cut of the correlation
functions for negative coupling.
They act as a second saddle point of the Euclidean
action. The (longitudinal) fluctuation operator around the 
saddle point has one negative eigenvalue,
commensurate with the imaginary 
square root of the determinant of this operator.

Through the evaluation of corrections 
to the classical configurations, 
we are able to evaluate corrections of 
relative order $g$ to the correlation functions,
which, via dispersion relations,
immediately lead to the corrections of 
relative order $1/K$ to the perturbative 
coefficients.
The connection is elucidated in great detail
in Sec.~\ref{sec12}.

In all cases, the leading term of the order-$g$
correction, in the large-$N$ limit,
is given by the multiplicative correction 
to the partition function term, i.e., due to the factor $\calF_\calZ$
given in Eq.~\eqref{defFZ}.
One might wonder why the term of order $N^2$ in the correction
to the partition function constitutes the 
universal leading correction in the large-$N$ limit.
In order to understand this phenomenon, let
us consider the computational origin of the corrections.
The leading term in a correlation function
(for the imaginary part) is given by a term
in which one replaces all field configurations by instantons.
Then, a set of universal corrections is obtained
when one keeps the instanton field configurations
inside the main integrand but considers the 
corrections due to the field Jacobian, and due to the
effective action around the instanton,
which, together, give rise to the 
universal correction factor $\calF_\calZ$ given in 
Eq.~\eqref{defFZ}.

Let us now consider the additional corrections
obtained when one replaces, instead, one of the 
instanton field configurations by a fluctuation.
Then, for the order-$g$ corrections to the 
imaginary part of the partition function,
one has to combine the fluctuation with the 
factor $\calF_\calJ$ from Eq.~\eqref{defFJ}.
This combination leads, at most, with regard
to $N$, to the product of a longitudinal fluctuation factor,
which carries no $N$, and one additional factor $N$
due to the transverse fluctuations encoded in 
$\und\chi^2 = \chi_\LL^2 + \und\chi_\TT^2$.
Finally, replacing two instanton configurations
by fluctuations, one obtains at most a single
factor of $N$, generated by a 
term proportional to $\und\chi^2$.
The universality of the large-$N$ limit 
of the correction terms can be justified based
on the decoupling of expectation values of 
fields at different space-time points in the 
limit of large $N$,
as explained in the text following Eq.~(2.2) 
of Ref.~\cite{MoZJ2003}
and Chap.~14 of Ref.~\cite{ZJ2007}.
In fact, according to formulas given in 
Chap.~14 of Ref.~\cite{ZJ2007},
critical exponents reach universal values
in the large-$N$ limit and depend only on the 
spatial dimension of the system.

Our expressions for the two-point correlation function, for its second
derivative, for the four-point correlation function 
and for the two-point correlation function with a wigglet insertion, computed at zero
momentum, were obtained by
integrating the corresponding correlation function in coordinate space over the
difference of all its coordinate with respect to one of them. 
Incidentally, it is interesting to note that this procedure is 
completely equivalent to an alternative procedure 
where one fixes one of the time coordinates in a correlation
function to zero, and integrates over all the others.
This equivalence holds due to the time translation invariance 
of the correlation functions.
One of the most important additional conclusions of the 
current article is that, for the correlation functions,
the two-loop corrections to the imaginary part, 
of relative order $g$, have a much more complex 
analytic structure as compared to those of the 
ground-state energy.

A remark on our notation is in order.
We apologize to the reader for the many $G$'s
in our paper. Generic $n$-point Green functions
are denoted by the sans-serif  $\sfG$,
according to Eq.~\eqref{sfGdef}. We also recall that 
$\calG_{\alpha\beta}$, according to Eq.~\eqref{defGab},
is the imaginary part of the two-point function, while
$\calG_{\alpha\beta\gamma \delta}$, according to Eq.~\eqref{defGabcd}, 
is the imaginary part of the four-point function.  Incidentally, 
we also have the terms $\calR_i$
($i = 1,\dots, 7$) for the contributions to the 
two-point function at zero momentum transfer, 
according to Eq.~\eqref{interim7}.
Seven corrections are also incurred 
for the contributions to the 
derivative of the two-point function,
summarized in the terms $\calS_i$
[see Eq.~\eqref{GGP_SEVEN}].
For the four-point function
one defines the scalar Green function
${\bfG}$ according to Eq.~\eqref{G4ZERODEF}.
Finally, we have the seven terms 
$\calT_i$ ($i = 1,\dots, 7$) 
for the four-point function
at zero momentum transfer, according to 
Eq.~\eqref{GGp_expr}, and the seven
terms $\calU_i$ ($i = 1,\dots, 7$) 
for the two-point function
with a wigglet insertion, according to Eq.~\eqref{G12RES}.

A remark on the character of the factorial 
divergence of the perturbation series 
is in order. According to Eq.~\eqref{genexp},
all factorially divergent series 
calculated in this work 
are alternating in large order, in
view of the factor $(-1/A)^K$ in  Eq.~\eqref{genexp},
where $A=4/3$.
Indeed, it is known that perturbation series
in $\phi^4$ theories are factorially
divergent, Borel summable series~\cite{Pa1978,BrPa1978reprinted,%
CaGrMa1986,BrLGZJ1977prd1,BrLGZJ1977prd2,ZJ2002}.
In a more general context, 
such series constitute the conceptually 
simplest manifestation of so-called 
resurgent expansions (transseries), which
have recently been shown to lead to adequate
representations of physical quantities 
of interest in a number of mathematical
investigations, and in field 
theories~\cite{Ph1989,DePh1998i,DePh1998ii,%
CaNoPh1993,JeZJ2004plb,ZJJe2004i,ZJJe2004ii,%
ResurgenceCERN}.

In principle, based on the results presented 
in the current paper, one could go further 
and calculate the large-order behavior of 
the renormalization constants of the one-dimensional
$O(N)$ field theory (see Sec.~\ref{appendixc}).
We recall that the renormalization constants
$Z_\phi$ (wave function),
$Z_{\phi^2}$ (wigglet insertion),
$Z_g$ (coupling constant), and
$\delta m^2$ (mass counter term),
are determined by the renormalization conditions
imposed on the vertex functions,
which can be obtained from the 
correlation functions (calculated here) 
via a Legendre transformation.
The renormalization constants, in turn, 
determine the large-order behavior of
the beta function $\beta(g)$,
the anomalous dimension function $\eta(g)$, and the
correlation length function $\eta_2(g)$,
which enter the Callan--Symanzik equation.
The Callan--Symanzik equation is a renormalization-group
(RG) equation fulfilled by the vertex functions of the theory
(a mini review on this aspect is given in Appendix~\ref{appendixc}).
However, in one dimension, we refrain from
engaging in this endeavor because 
of the absence of a second-order phase transition
due to the low dimensionality of the system under
study, which prevents the system from undergoing a 
phase transition to the low-temperature phase.
In higher dimensions, the critical exponents can
be studied on the basis of the Callan--Symanzik 
equation~\cite{LGZJ1977,LGZJ1980,LGZJ1990,GuZJ1998}.

One of the main conclusions of the 
current paper is that the calculation 
of corrections to the large-order 
growth of perturbation theory 
for correlation functions, beyond the plain
calculation of the partition function,
is possible for field theories with a 
nontrivial internal structure (here, the 
$O(N)$ symmetry group).
The results presented here are a first step
toward the evaluation of subleading 
corrections to the factorial asymptotics of 
perturbative coefficients in perturbative
field theory, for physical quantities of
interest beyond the partition function.
The evaluation of subleading corrections 
to the large-order asymptotics of Feynman diagram
coefficients in large loop order provides us 
with an alternative method to enhance our
understanding of the predictive limits of field theory.
The ultimate goal of the endeavor 
is to ``interpolate'' between (necessarily finite-order)
perturbative Feynman diagram calculations and 
the large-order asymptotics (about ``infinite loop'' order),
the latter being enhanced by the evaluation of the 
subleading corrections about the instantons.
The latter, in turn, lead to corrections 
of relative order $1/K$ (two-loop order discussed here)
to the large-order behavior of
perturbation theory.
The generalization
to relative order $1/K^2$ (four-loop order about the instanton)
and the consideration of field theories in higher 
dimensions are the natural next steps in this program
and currently under investigation.

\section*{Acknowledgements}

This work has been supported by the National 
Science Foundation (Grant PHY--1710856)
and by the Swedish Research Council (Grant No.~638-2013-9243). 
Support from the Simons Foundation
(Grant 454949) also is gratefully acknowledged.

\appendix

\renewcommand\theequation{\thesection.\arabic{equation}}

%
%
\section{Integral Table}
\label{appendixb}

We first list integrals that only involve the instanton,
\begin{align} 
\label{intH1}
H_1 = & \; \int \dd t \, \xi_\cl(t) = \sqrt{2} \, \pi \,,
\\[0.1133ex]
\label{intH2}
H_2 = & \; \int \dd t \, [\xi_\cl(t)]^2 = 4 \,,
\\[0.1133ex]
\label{intH3}
H_3 = & \; \int \dd t \, t^2 \, \xi_\cl(t) = \frac{\pi^3}{2 \sqrt{2}} \,.
\end{align}
Integrals involving the subtracted longitudinal and 
transverse propagators read as follows,
\begin{align}
\label{I1}
I_1 =& \; 
\int \dd t_1 \, \int \dd t_2 \, [\bfDelta_\LL(t_1, t_2) - \bfDelta_0(t_1, t_2)] = - 4 \,
\\[0.1133ex]
\label{I2}
I_2 =& \; 
\int \dd t_1 \, \int \dd t_2 \, [\bfDelta_\TT(t_1, t_2) - \bfDelta_0(t_1, t_2)] 
\nonumber\\[0.1133ex]
=& \; 1 - \frac{\pi^2}{4} - \frac72 \, \zeta(3) \,.
\end{align}
In addition, we have
integrals involving longitudinal and transverse propagators
and instantons,
\begin{align}
\label{J1}
J_1 =& \; \int \dd t_1 \, \xi_\cl(t_1) \, \int \dd t_2 \, 
\bfDelta_\LL(t_1, t_2) = 0 \,,
\\[0.1133ex]
\label{J2}
J_2 =& \; \int \dd t_1 \, \ddot\xi_\cl(t_1) \, \int \dd t_2 \, 
\bfDelta_\LL(t_1, t_2) 
= \frac{\pi}{\sqrt{2}} \,,
\\[0.1133ex]
\label{J3}
J_3 =& \; \int \dd t_1 \, \xi_\cl(t_1) \, \bfDelta_\LL(t_1, t_1) \, 
\int \dd t_2 \, \bfDelta_\LL(t_1, t_2) 
= \frac{\pi}{4 \sqrt{2}} \,,
\\[0.1133ex]
\label{J4}
J_4 =& \; \int \dd t_1 \, \xi_\cl(t_1) \, \bfDelta_\TT(t_1, t_1) \, 
\int \dd t_2 \, \bfDelta_\LL(t_1, t_2) 
= \frac{\pi}{8 \sqrt{2}} \,.
\end{align}
Integrals involving the subtracted longitudinal and
transverse propagators, and powers of the 
Euclidean time variable,
are needed for the second derivative of the two-point function,
\begin{align}
\label{K1}
K_1 =& \; \int \dd t_1 \, \int \dd t_2 \, t_1^2 \,
[\bfDelta_\LL(t_1, t_2) - \bfDelta_0(t_1, t_2)] 
\nonumber\\[0.1133ex]
=& \; -\frac{\pi^2}{3} \,,
\\[0.1133ex]
\label{K2}
K_2 =& \; \int \dd t_1 \, \int \dd t_2 \, t_1^2 \,
[\bfDelta_\TT(t_1, t_2) - \bfDelta_0(t_1, t_2)] 
\nonumber\\[0.1133ex]
=& \; 2 + \frac{\pi^2}{12} - \frac{\pi^4}{16} 
- \frac{7}{24} \pi^2 \zeta(3) - \frac{31}{2} \zeta(5) \,.
\end{align}
Furthermore, we have the following integrals with 
mixed contributions,
\begin{align}
\label{K1tilde}
{\widetilde K}_1 =& \; \int \dd t_1 \, \int \dd t_2 \, t_1 \, t_2 \,
[\bfDelta_\LL(t_1, t_2) - \bfDelta_0(t_1, t_2)] 
\nonumber\\[0.1133ex]
=& \; -\frac{7 \pi^2}{12} - 1 - \frac{21}{2} \zeta(3) \,,
\\[0.1133ex]
\label{K2tilde}
{\widetilde K}_2 =& \; \int \dd t_1 \, \int \dd t_2 \, t_1 \, t_2 \,
[\bfDelta_\TT(t_1, t_2) - \bfDelta_0(t_1, t_2)] 
\nonumber\\[0.1133ex]
=& \; -1 + \frac{\pi^2}{12} - \frac{7}{24} \pi^2 \zeta(3) + \frac{31}{4} \zeta(5) \,.
\end{align}

Among all integrals considered, the integrals $K_2$ and 
${\widetilde K}_2$ are 
by far the most difficult to evaluate.

One may derive a relatively compact representation for 
$K_2$ via the substitutions $t_2 = 2 \, \ln(v)$, 
followed by the $t_2$ integral, then followed 
by $t_1 = 2 \, \ln(u)$, and $u = \sqrt{x}$, 
so one would effectively substitute $t_1 =  \ln(x)$.
This results in 
\begin{align}
K_2 =& \; \int_0^\infty \dd x \, 
\left\{ -\frac{\ln^2(x)}{x (x^2 + 1)}
- \frac{\ln^2(x) \, (x^2 - 1) \, \arctan(x)}{x^2 \, (x^2 + 1)}
\right.
\nonumber\\[0.1133ex]
& \; \left. - \frac{\ii \ln^2(x)}{1 + x^2} \, \left[
{\LLi}_2(-\ii x) - {\LLi}_2(\ii x) \right] 
\right\} \,.
\end{align}
This representation is seen to involve
Legendre's $\chi_2$ function~\cite{Le1981,Le1811},
\begin{equation}
\chi_2(z) = \frac12 \,
\left( {\LLi}_2(z) - {\LLi}_2(-z) \right) \,.
\end{equation}
Formula~\eqref{K2} was found by the 
PSLQ algorithm~\cite{FeBa1992,BaPl1997,FeBaAr1999,BaBr2001}.
One can form the combinations,
\begin{align}
\label{M1}
{\overline K}_1 =& \; \int \dd t_1 \, \int \dd t_2 \, (t_1 - t_2)^2 \,
[\bfDelta_\LL(t_1, t_2) - \bfDelta_0(t_1, t_2)] 
\nonumber\\[0.1133ex]
=& \; 2 (K_1 - {\widetilde K}_1) 
= 2 + \frac{\pi^2}{2} + 21 \zeta(3) \,,
\\[0.1133ex]
\label{M2}
{\overline K}_2 =& \; \int \dd t_1 \, \int \dd t_2 \, (t_1 - t_2)^2 \,
[\bfDelta_\TT(t_1, t_2) - \bfDelta_0(t_1, t_2)] 
\nonumber\\[0.1133ex]
=& \; 2 (K_2 - {\widetilde K}_2) 
= 6 - \frac{\pi^4}{8} - \frac{93}{2} \zeta(5) \,.
\end{align}

We have additional reference integrals with a 
second derivative insertion, 
which are important for the calculation of the derivative
$\partial^2 G/\partial p^2$, at $p=0$,
\begin{align}
\label{L1}
L_1 =& \; \int \dd t_1 \, \xi_\cl(t_1) \, \int \dd t_2 \, 
t_2^2 \, \bfDelta_\LL(t_1, t_2) = 
\frac{\pi^3}{2 \sqrt{2}} \,,
\\[0.1133ex]
\label{L2}
L_2 =& \; \int \dd t_1 \, \ddot\xi_\cl(t_1) \, 
\int \dd t_2 \, t_2^2 \, \bfDelta_\LL(t_1, t_2) 
= \frac{3 \pi^3}{4 \sqrt{2}} \,,
\\[0.1133ex]
\label{L3}
L_3 =& \; \int \dd t_1 \, \xi_\cl(t_1) \, \bfDelta_\LL(t_1, t_1) \, 
\int \dd t_2 \, t_2^2 \, \bfDelta_\LL(t_1, t_2) 
\nonumber\\[0.1133ex]
=& \; \frac{5 \pi^3}{16 \sqrt{2}} - \frac{\pi}{12 \sqrt{2}} \,,
\\[0.1133ex]
\label{L4}
L_4 =& \; \int \dd t_1 \, \xi_\cl(t_1) \, \bfDelta_\TT(t_1, t_1) \, 
\int \dd t_2 \, t_2^2 \, \bfDelta_\LL(t_1, t_2) 
\nonumber\\[0.1133ex]
=& \; \frac{9 \pi^3}{32 \sqrt{2}} \,.
\end{align}
For the wigglet insertion, i.e., the Green function
$ \calG^{(1,2)} $ at zero momentum, we also need the integrals
\begin{align}
M_1 = & \; 
\int \dd t_1 \, \xi_{\rm cl}(t_1)\, 
\int \dd t_2 \, 
\xi_{\rm cl}(t_2) \, \bfDelta_{\LL}(t_1,t_2) 
= -1  \,,\\
M_2 = & \; \int \dd t_1 \, {\ddot \xi}_{\rm cl}(t_1) \, 
\int \dd t_2 \, \xi_{\rm cl}(t_2)\, \bfDelta_{\LL}(t_1,t_2) 
= 1 \,,
\\[0.1133ex]
M_3 = & \; 
\int \dd t_1 \, 
\xi_{\rm cl}(t_1) \, 
\bfDelta_{\LL}(t_1, t_1) \, 
\int \dd t_2 \, 
\xi_{\rm cl}(t_2) \, 
\bfDelta_{\LL}(t_1,t_2) 
\nonumber\\[0.1133ex]
=& \; \dfrac{1}{36} \,,
\\[0.1133ex]
M_4 = & \; 
\int \dd t_1 \, \xi_{\rm cl}(t_1) \, \bfDelta_{\TT}(t_1, t_1) \, 
\int \dd t_2 \, \xi_{\rm cl}(t_2) \, 
\bfDelta_{\LL}(t_1,t_2) 
\nonumber\\[0.1133ex]
=& \; -\dfrac{1}{4} \,.
\end{align}
In view of the confluence of arguments of the 
propagators in the wigglet insertion, additional integrals are 
required,
\begin{align}
N_1 =& \; \int \dd t_1 \,
\left[ \bfDelta_{\LL}(t_1, t_1) - \bfDelta_0(t_1, t_1) \right] = -\dfrac{7}{6}\,,\\
N_2 =& \; \int \dd t_1 \, \,
\left[ \bfDelta_{\TT}(t_1, t_1) - \bfDelta_0(t_1, t_1) \right] =\,-\dfrac{1}{2} \,.
\end{align}
Finally, we also need the integral
\begin{equation}
J_1^{(0)} = 
\int \dd t_1 \, \xi_\cl(t_1)
\int \dd t_2 \, \bfDelta_0(t_1, t_2) = \sqrt{2} \, \pi \,,
\end{equation}
which is obtained from~\eqref{J1} by the replacement
of the longitudinal propagator with its free counterpart.

%
%
\section{Callan--Symanzik Equation}
\label{appendixc}

In principle, the application
of the Callan--Symanzik equation~\cite{Ca1970,Sy1970}
to the calculation of critical exponents has
been described in a number of 
monographs~\cite{Pa1988stat,ZJ2002,ZJ2007,KlSF2007}.
However, in order to put the calculations 
reported here into perspective, 
we should include a few remarks regarding the 
relation of the quantities calculated
here, to the renormalization-group (RG) functions
that enter the Callan--Symanzik equation.
The Callan--Symanzik equation is a 
renormalization-group (RG) equation fulfilled
by one-particle irreducible vertex functions
$\Gamma^{(a,n)}$, derived via Legendre
transformation from the connected
correlation functions (see Chap.~7 of Ref.~\cite{ZJ2002}).

The Callan--Symanzik equation is obtained by differentiating
the vertex functions with respect to the 
renormalized mass parameter while
holding the bare parameters constant, 
and reads as follows,
\begin{multline}
\label{Callan--Symanzikeq}
\biggl[ m_R \frac{\partial}{\partial m_R} +
\beta(g_R) \, \frac{\partial}{\partial g_R} -
\frac{n}{2} \eta(g_R)
- a \eta_2(g_R) \biggr) \biggr] 
\\[0.1133ex]
\Gamma_R^{(a,n)}\left(
\left\{ \vec q_i \right\}_{i=1}^a; 
\left\{ \vec p_i \right\}_{i=1}^n; m_R, g_R\right)
\\[2ex]
= \left[ 2 - \eta(g_R) \right] \Gamma_R^{(a+1,n)}\left(
\left\{ \vec q_i \right\}_{i=1}^a, \vec 0;
\left\{ \vec p_i \right\}_{i=1}^n;
m_R, g_R \right) \,.
\end{multline}
Here, $\Gamma_R^{(a,n)}$ is the $n$-point vertex 
function with $a$ wigglet insertions.
In the term with $\Gamma_R^{(a+1,n)}$, we have
to set the momentum argument corresponding to the 
wigglet insertion number $a+1$ [the one created 
by the action of the operator $\partial/\partial m_R^2$ 
on $\Gamma_R^{(a,n)}$] to zero.
As already mentioned, the vertex functions are obtained from the 
connected correlation functions via a Legendre transformation.
For example, the vertex function 
$\Gamma^{(2)} = \Gamma^{(0,2)}$ is the 
inverse of the two-point correlation function $W^{(2)}$ 
[see the discussion following Eq.~(7.80) of Ref.~\cite{ZJ2002}].
The four-point function $\Gamma^{(4)} = \Gamma^{(0,4)}$ is 
obtained from $W^{(4)}$ via ``amputation'' of the 
external legs, and sign inversion
[see the discussion following Eq.~(7.80) of Ref.~\cite{ZJ2002},
or Eq.~(4.24) of Ref.~\cite{KlSF2007}].

Bare (index zero) and renormalized vertex functions are related by
[see Eq.~(10.20) of Ref.~\cite{ZJ2002}]
\begin{multline}
\label{ZphiZphi2}
\Gamma_0^{(a,n)}\left(
\left\{ \vec q_i \right\}_{i=1}^a;
\left\{ \vec p_i \right\}_{i=1}^n;
m_0, g_0 \right) \\
= Z_\phi^{-n/2} \, \left( \frac{Z_\phi}{Z_{\phi^2}} \right)^a \,
\Gamma_R^{(a,n)}\left(
\left\{ \vec q_i \right\}_{i=1}^a;
\left\{ \vec p_i \right\}_{i=1}^n;
m_R, g_R \right) \,.
\end{multline}
Here, $Z_\phi$ is the wave function renormalization, 
$Z_{\phi^2}$ is the renormalization of the wigglet 
insertion, and the bare and renormalized 
mass parameters are related by 
\begin{equation}
m_0^2 = (m_R^2 + \delta m^2)/Z_\phi \,,
\end{equation}
where $\delta m^2$ is the mass counter term.

The wave function renormalization constant $Z_{\phi^2}$
is fixed by the condition
\begin{equation}
\label{ren_wave1}
\left. \frac{\partial}{\partial p_1^2} 
\Gamma^{(2)}_R(\vec p_1, \vec p_2; m^2_R, g_R) 
\right|_{\vec p_i = \vec 0} = 1 \,.
\end{equation}
The renormalization condition
\label{RGcond}
\begin{equation}
\label{ren_mass}
\left.
\Gamma^{(2)}_R(\vec p_1, \vec p_2; m_R^2, g_R) 
\right|_{\vec p_i = \vec 0} = m_R^2
\end{equation}
fixes the mass counter term $\delta m^2$.
The coupling constant renormalization constant $Z_g$ is 
fixed by the condition
\begin{equation}
\label{ren_gR}
\left. \Gamma^{(4)}_R(\vec p_1, \vec p_2; \vec p_3, \vec p_4,
m^2_R, g_R) \right|_{\vec p_i = \vec 0} = m_R^{4-D} \; g_R \,,
\end{equation}
where $g_R$ is the dimensionless, renormalized coupling,
and $D$ is the spatial dimension.
Finally, the wigglet insertion renormalization constant
$Z_{\phi^2}$ is determined by 
\begin{equation}
\label{ren_wave2}
\left.  \Gamma^{(1,2)}_R( \vec q; \vec p_1, \vec p_2; m_R^2, g_R) 
\right|_{\vec p_i, \vec q = 0} = 1 \,.
\end{equation}
The renormalization-group functions 
are obtained as follows,
\begin{align}
\beta(g_R) =& \;
- (4-D) \left[ \frac{\partial}{\partial g_R} \ln\left( g_R \,
\frac{Z_g(g_R)}{Z_\phi(g_R)^2} \right) \right]^{-1} \,,
\\[0.1133ex]
\eta(g) =& \; 
\beta(g_R) \, \frac{\partial}{\partial g_R} 
\ln( Z_\phi ) \,,
\\[0.1133ex]
\eta_2(g) =& \; 
\beta(g_R) \, \frac{\partial }{\partial g_R} 
\ln\left( \frac{Z_{\phi^2}}{Z_\phi} \right) \,.
\end{align}
From these relations, one calculates the 
critical value $g^*$ determined by the condition
$\beta(g^*)=0$ which determines the infrared non-Gaussian
fixed point of the RG flow, 
which is approached by the system because of the 
relevance of the $\phi^4$ interaction in dimensions
lower than four. 
Critical exponents are determined by the values of the 
RG functions at the critical point,
$\eta = \eta(g^*)$ and $\eta_2 = \eta_2(g^*)$,
via well-known hyperscaling relations.
For example, the critical exponent $\nu$ of the correlation
length is determined as $\nu = 1/(2 + \eta_2)$,
while the critical exponent $\alpha$ of the 
heat capacity is $\alpha = 2 - D \, \nu$,
and the critical exponent $\gamma$ of the magnetic 
susceptibility is
$\gamma = \nu \, (2 - \eta)$.

\end{document}